\documentclass{aa}   
\usepackage{graphicx}
\usepackage[varg]{txfonts}
\usepackage{longtable}
\usepackage{natbib}
\usepackage{url}
\usepackage{color}
\usepackage{multirow,bigdelim}
\usepackage{cases}
\usepackage{rotating}
\usepackage[colorinlistoftodos]{todonotes}
\bibpunct{(}{)}{;}{a}{}{,} 
\newcommand\kms{{\rm\,km\,s^{-1}}}

\usepackage{hyperref} 
\hypersetup{
     breaklinks,
     colorlinks   = true,
     citecolor    = blue,
     urlcolor     = blue,
     linkcolor    = blue
}

\begin{document} 

\title{Searching for and characterizing halo substructures with the GALAH DR4 survey}

\titlerunning{Searching for and characterizing halo substructures with the GALAH DR4 survey} 

\author{
Iryna Kushniruk\inst{1}
\and
Kristopher Youakim\inst{1}
\and 
Karin Lind\inst{1}
\and 
Sven Buder\inst{2,6}
\and 
Janes Kos\inst{4}
\and
Diane Feuillet\inst{8}
\and
Sarah L. Martell\inst{6,7}
\and
Richard de Grijs\inst{5,9,10}
\and
Geraint F. Lewis\inst{3}
\and
Joss Bland-Hawthorn\inst{3,6}
\and 
Gary Da Costa\inst{2}
\and  
Michael Hayden\inst{3,6}
\and
Daniel Zucker\inst{5,6}
\and 
Tomaz Zwitter\inst{4}
\and 
Sanjib Sharma\inst{11}
}

\institute{
$^{1}$
Department of Astronomy, Stockholm University, AlbaNova University Center, SE-106 91 Stockholm - Sweden \\
$^{2}$
Research School of Astronomy and Astrophysics, The Australian National University, Canberra ACT2611, Australia\\
$^{3}$
Sydney Institute for Astronomy, School of Physics, A28, The University of Sydney, NSW 2006, Australia\\
$^{4}$
Faculty of Mathematics and Physics, University of Ljubljana, Jadranska 19, 1000 Ljubljana, Slovenia\\
$^{5}$
School of Mathematical and Physical Sciences, Macquarie University, Balaclava Road, Sydney, NSW 2109, Australia\\
$^{6}$
ARC Centre of Excellence for All Sky Astrophysics in 3 Dimensions (ASTRO 3D), Australia\\
$^{7}$
School of Physics, University of New South Wales, Sydney NSW 2052, Australia\\
$^{8}$
Lund Observatory, Department of Geology, Sölvegatan 12, SE-223 62 Lund, Sweden\\
$^{9}$
Astrophysics and Space Technologies Research Centre, Macquarie University, Balaclava Road, Sydney, NSW 2109, Australia\\
$^{10}$
International Space Science Institute--Beijing, 1 Nanertiao, Zhongguancun, Hai Dian District, Beijing 100190, China\\
$^{11}$
Space Telescope Science Institute, 3700 San Martin Drive, Baltimore, MD 21218, USA\\
\email{iryna.kushniruk@astro.su.se}
}

\date{Received 20 June 2024 / Accepted 09 Feb 2026}

 
\abstract
{
Recent studies have revealed that the Milky Way's stellar halo is a composite of stellar populations of different origins, including multiple accretion events. To better understand how the Milky Way and other spiral galaxies were formed, it is necessary to thoroughly characterize the chemical and kinematic properties of these structures.
}
{
We search for kinematic structures of the stellar halo, find any substructures within them if present, and characterize the chemo-dynamical properties of the identified groups with the GALAH DR4 and {\it Gaia} surveys.
}
{
We apply wavelet transforms in the space defined by a square root of radial action ($J_r$) and azimuthal action ($L_z$) to search for kinematic overdensities. Then, we select stars in the detected structures and investigate their elemental abundances to determine their origin. Additionally, we check for any contamination from other stellar populations within the detected groups with the unsupervised machine learning algorithm t-Distributed Stochastic Neighbor Embedding (t-SNE), for which we perform chemical tagging in a high-dimensional parameter space using 15 elemental abundances as input.
}
{
We recovered five kinematic structures in the action space with the wavelet transform. These groups are the Galactic disk, Splash, {\it Gaia}-Sausage-Enceladus (GSE), Thamnos1 and Thamnos2. We found that GSE has two peaks with the wavelet transform; one of these peaks is located at $\sqrt{J_r} \simeq 25$ kpc $\kms$ and is a result of contamination from disk stars. The other peak corresponds to the 'cleanest' GSE population and is located above $\sqrt{J_r} \simeq 40$ kpc $\kms$. We also detected three peaks in Thamnos. We linked two of them to Thamnos1 and the peak with the stars on the most retrograde orbits to Thamnos2. The t-SNE algorithm also confirmed these findings.  We also analyzed individual elemental abundances of each group and found that Thamnos2 has a higher [$\alpha$/Fe] ratio than the other groups, iron-peak elements are more abundant in the Splash than in the halo groups, while the halo structures retain a higher r-process signature than the splashed disk.  
}
{ 
A multiply-peaked substructure we observe in action space in GSE and Thamnos suggests that the splashed disk extends beyond the borders of prograde orbits. Each of the four halo groups studied in this paper have unique chemo-dynamical properties that confirm their extra-galactic origin.
}


\maketitle
%

\section{Introduction}
The formation and evolution process of large spiral galaxies is an active research area in modern astronomy. Since the Milky Way is the only galaxy whose stars and structures can be studied in great detail, it is a benchmark for constraining models of galaxy formation. Therefore, studying the formation of the Milky Way in great detail is crucial to our understanding of spiral galaxies in general.

It is known that the Galaxy contains many kinematic structures, and the nature and origins of these remain unresolved. A vast number of currently known structures have been found in the Galactic halo and likely formed due to accretion events where small galaxies merged with the Milky Way \citep[e.g.][]{_helmi18, _belokurov18, _koppelman19, _myeong19, _helmi20, _naidy20, _s5_20, _deason24}. Since accreted stars retain similar chemical composition and motion to their progenitor \citep[e.g.][]{_freeman02, _tolstoy09}, examining the chemo-dynamical properties of stars helps to separate accreted populations from those formed {\it in situ}. Thus, we need to have high-precision spectroscopic parameters and astrometric measurements for as many stars in the Galaxy as possible to discover new halo populations and characterize the already known ones. 

Nowadays, large spectroscopic and astrometric surveys that provide data for millions of stars, and many millions more are expected in the near future. Current and future surveys include the GAlactic Archaeology with HERMES \citep[GALAH;][]{_desilva15}, the Apache Point Observatory Galactic Evolution Experiment \citep[APOGEE;][]{_holzman18}, the Large sky Area Multi-Object fibre Spectroscopic Telescope \citep[LAMOST;][]{_zhao12}, the Hectochelle in the Halo at High resolution survey \citep[H3;][]{_conroy19}, {\it Gaia} \citep{_lindegren16} and upcoming surveys like the 4-metre Multi-Object Spectroscopic Telescope \citep[4MOST;][]{_dejong2019} and the WHT Enhanced Area Velocity Explorer \citep[WEAVE;][]{_dalton18, _dalton20} are expected to provide data on many more millions of stars. These and many other surveys provide the necessary data to explore halo debris, which has led to several discoveries. For instance, {\it Gaia}-Sausage-Enceladus (GSE) \citep{_belokurov18, _helmi18}, which is linked to the last major merger that happened roughly 10 Gyr ago; the Sequoia accretion event \citep{_myeong19}; Thamnos \citep{_koppelman18}; Heracles, the accreted population in the inner disk \citep{_horta21}. It has also been shown that major merger events like GSE can perturb the Galactic disk \citep{_belokurov20} and lead to the formation of kinematic structures like the Splash. All of these recent findings raise interest in investigating and characterizing the structures further.

The discovery of these halo substructures has raised the question of how one should select stars that come from the accreted progenitor only and avoid contamination by halo stars from other sources like the Galactic disk. For instance, \citet{_feuillet20} found that to get a "clean" subset of GSE stars, one should select stars with a square root of radial actions $\sqrt{J_r}$ > 30 kpc $\kms$. A study by \citet{_buder22} shows that kinematically, GSE extends towards lower values of a square root of a radial action with a mean value around $\sqrt{J_r}\sim$ 26 kpc $\kms$ and less than 30\% of stars stay within the "clean" area. This means that purely kinematic criteria would remove a significant number of GSE stars with a lower square root of radial actions below 30 kpc $\kms$. Later, \citet{_feuillet21} analyzed stars from the APOGEE and {\it Gaia} surveys and found that kinematically selected GSE and Sequoia populations contain a small number of {\it in situ} stars, which were identified after examining their elemental abundances.

Accreted populations stand out from the disk stars not only dynamically but also chemically, and most of them occupy partially overlapping regions in a multidimensional abundance space. For example, a study by \citet{_contursi23} suggested that GSE has a higher Ce abundance than the Helmi streams, which is also an accreted population discovered by \citet{_helmi99}. It is known that GSE stars have lower [$\alpha$/Fe] ratios compared to other halo stars, suggesting a different formation history \citep[e.g.][]{_haywood18}. Other structures, like Thamnos and Heracles, exhibit a higher [$\alpha$/Fe] than GSE \citep{_horta23}. In addition, \citet{_matsuno21} found that GSE is enhanced in r-process elements while being under-abundant in s-process elements. The mass estimate of the GSE progenitor by \citet{_lane23} is $\sim1.45\times10^8 M_{\odot}$ and is roughly 15-25\% of the Milky Way's stellar halo mass. The stellar mass of Thamnos is estimated to be $\sim5\times10^6 M_{\odot}$ \citep{_koppelman19}. The GSE accretion event occurred around 8-9 billion years ago, according to \citet{_helmi18}, while some recent studies, for instance, \citet{_feuillet21} suggest a slightly older age of 10-12 Gyr. At the time, tidal interactions between the disk and the GSE progenitor caused some of the old disk stars to gain vertical energy during the accretion event, creating a kinematically hot and physically thick disk with the same abundance pattern as the pre-interaction disk. According to \citet{_belokurov20}, these stars formed a kinematic structure known as the Splash.

Although so much progress has been made recently in characterizing these halo structures, many questions still need to be answered. One is whether there is any substructure within these newly discovered groups, especially in the GSE, as it is one of the largest known accreted groups to date. Answering this question would help us better understand the past and the structure of the GSE progenitor and our Galaxy. In this paper, we aim to search for kinematic structures of the stellar halo and any substructure within them and characterize their chemo-dynamical properties with the fourth data release from the GALAH survey \citep{_galah_dr4}.

We structure the paper in the following way. In Sect.~\ref{_sec_data}, we describe the surveys used in this work and the construction of the stellar sample. In Sect.~\ref{_sec_analysis_res}, we explain the methodology we used to search for structures in the stellar halo and present the main results of this work. In Sect.~\ref{_sec_contamination}, we check how many disk stars contaminate the halo structures we identified. In Sect.~\ref{_sec_characterization}, we discuss the chemo-dynamical properties of each group. In Sect.~\ref{_sec_tsne}, we perform the analysis of our sample with an alternative method. We summarize the paper in Sect.~\ref{_sec_summary}.   

\section{Data}\label{_sec_data}
To search for and characterize substructures in the stellar halo, we need a combination of kinematic parameters and elemental abundances for thousands of stars covering a wide range of Galactocentric radii and heights above and below the plane of the Galactic disk. 

In this work, we use elemental abundances and radial velocities from the GALAH survey \citep{_desilva15}. GALAH is a high-resolution (R$\sim28\,000$) spectroscopic survey that provides stellar parameters and elemental abundances for stars in the Southern Hemisphere. We use the GALAH Data Release 4 (hereafter GALAH DR4; \citet{_galah_dr4}), which contains almost a million stars. We combine spectroscopic data from GALAH DR4 with coordinates, parallaxes, and proper motions from the {\it Gaia} mission and use its latest Data Release 3 \citep[hereafter {\it Gaia} DR3;][]{_gaiadr3}. The distance estimates were taken from a complementary catalog by \citet{_bailerjones23}.

\begin{figure}
   \centering
   \resizebox{\hsize}{!}{
   \includegraphics[viewport = 0  0 520 350,clip]{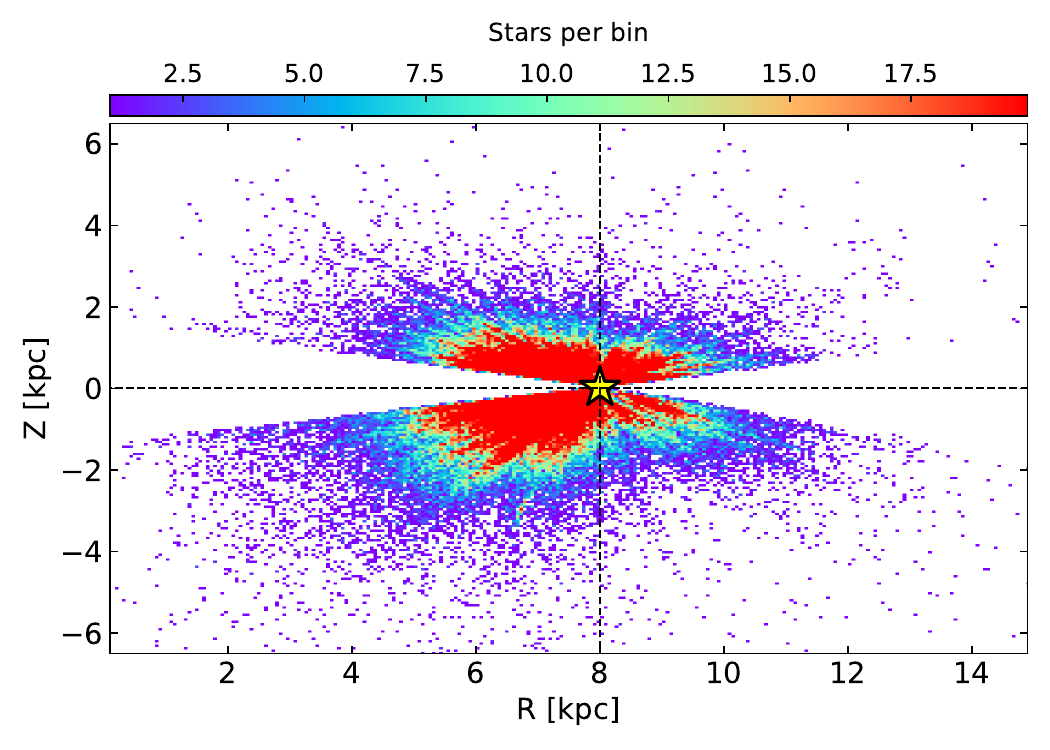}
   }
   \caption{Galactocentric distance, R, as a function of distance from the Galactic plane, Z, in the direction of the North Galactic Pole, Z $=90^{\circ}$, for 124\,618 stars selected from GALAH DR4. Dashed lines show the Solar values R $=8$ kpc, and Z $=20.8$ pc and the yellow star shows the location of the Sun. The bin size is $0.05\times0.05$ kpc.
   \label{_fig_rz}
   }
\end{figure}

To select the most reliable elemental abundances and astrometric data for the analysis, we apply the following cuts: 

\begin{enumerate}
\item \begin{verbatim}flag_sp == 0\end{verbatim}
\item \begin{verbatim}snr_px_ccd3 > 30\end{verbatim}
\item \begin{verbatim}ruwe < 1.4\end{verbatim}
\item 
\begin{minipage}[t]{0.9\linewidth}
\footnotesize
\begin{verbatim}
survey_name = 'galah_main' or
              'galah_bright' or 'galah_faint'
\end{verbatim}
\end{minipage}
\item \begin{verbatim}log g < 3.5\end{verbatim}
\item \begin{verbatim}4000 [K] < teff < 6500 [K]\end{verbatim}
\end{enumerate}

These cuts produce the best-quality spectroscopic (cuts 1-2) and astrometric data (cut 3) and exclude open and globular clusters and other sub-surveys (cut 4). By allying cut 1 we select stars with no identified problems with stellar parameter determination. With cut 2 we ensure to get stars with a high signal-to-noise ratio per pixel for CCD3. Cut 3 is a re-normalized unit weight error for stars and is taken from the {\it Gaia} DR3 catalog. To avoid selection effects that might arise while mixing different types of stars at a wide range of Galactocentric radii, we limited our study to red giants only (cuts 5-6). Due to their long lifetime and brightness, red giants are easily observed at great distances from the Sun, allowing us to explore different parts of the Galaxy, such as the stellar halo. 

As a result, we end up with a sample of 124\,618 stars. Figure~\ref{_fig_rz} shows the location of the selected stars in the Galactocentric cylindrical coordinate system. The Solar values of R $=8$~kpc and Z $=20.8$~pc and the Solar motion corrections $U_{\odot}=11.1 \kms$, $V_{\odot}=12.24 \kms$, and $V_{\odot}=7.25 \kms$ were taken from \citet{_galpy} and \citet{_z0}. The majority of stars can be found in the direction opposite the North Galactic Pole (NGP) and are predominantly located towards the inner Galaxy (R $\lesssim8$ kpc). GALAH DR4 covers a broad range of Galactocentric radii and heights from the plane, allowing us to search for kinematic structures in the local stellar halo. 

\section{Analysis and Results}\label{_sec_analysis_res}

We search for kinematic structures and examine a density distribution in the $L_z - \sqrt{J_r}$ space as was used in, for example, \citet{_trick18}. We applied the wavelet transform with the {\it\`{a} trous} algorithm, which is an advanced statistical method described in \citet{_starck06}. This method is commonly used to search for overdensities, including kinematic structures \citep[e.g.][]{_antoja12, _kushniruk17, _ramos18}. The wavelet transform has several significant advantages: no assumptions on the initial velocity distribution of stars are needed; the method allows us to detect a substructure in the sample if present at different decomposition levels proportional to the size of the structures we want to reveal. 

We calculate orbital actions such as angular momenta $L_z$ and the square root of radial actions $J_r$ for all stars, assuming that the Galactic potential is axisymmetric. These quantities are fully conserved in axisymmetric systems and thus can be used to search for kinematic structures \citep{_trick18}. The square root of $J_r$
is proportional to the oscillation amplitude in the radial direction, representing the star's epicyclic motion around its guiding radius. This relation is well illustrated in Figure 1 from \citet{_blandhawthorn21}. Therefore, using of $\sqrt{J_r}$ instead of $J_r$ is beneficial for dynamical characterization of stellar populations. To calculate actions we used the {\tt galpy}\footnote{\url{http://github.com/jobovy/galpy}} \citep{_galpy} package and the {\tt MWPotential2014} potential. The potential consists of the Navarro-Frenk-White halo, Miyamoto-Nagai disk, and a power-law bulge available in {\tt galpy}.  

To perform the wavelet transform, we use the {\tt PyWavelets}\footnote{\url{https://pywavelets.readthedocs.io}} package \citep{_pywt}. Wavelet coefficients are the main characteristic of the substructure and are proportional to the intensity of their detection. The input data is a binned 2D distribution. After several tests, we found that the optimal number of bins which could provide a good balance between resolution and noise suppression for our sample was 512. The most suitable wavelet function was {\it \mbox{Symlet 2}}, as it is symmetrical and has a minimal oscillation. Among other wavelet functions, we found that Daubechies wavelet {\it \mbox{db 2}} also recovered dynamical structures well in the data, but we chose {\it \mbox{Symlet 2}} based on its highly symmetrical properties. While probing a range of decomposition steps, $J$, from 2 to 6, we set this parameter to five and six as it gives the best visual recovery of dynamical groups. The relation between decomposition levels, bin size, and sizes of the structures we search for is $S_J = 2^{J}\Delta$\ \citep[for more details, see][]{_starck06}. In our case the bin size, $\Delta$, for 512 bins is [$\sqrt{J_r}, L_z$] $\simeq$ [0.2, 11.7] kpc~$\kms$. Thus, for scale $J=6$, we search for structures with sizes [$\sqrt{J_r}, L_z$] $\simeq$ [12.8, 748] kpc~$\kms$, and similarly for the scale $J=5$ we recover structures with sizes [$\sqrt{J_r}, L_z$] $\simeq$ [6.4, 374] kpc $\kms$. 

\begin{figure*}
   \centering
   \resizebox{\hsize}{!}{
   \includegraphics[viewport = 0  0 520 350,clip]{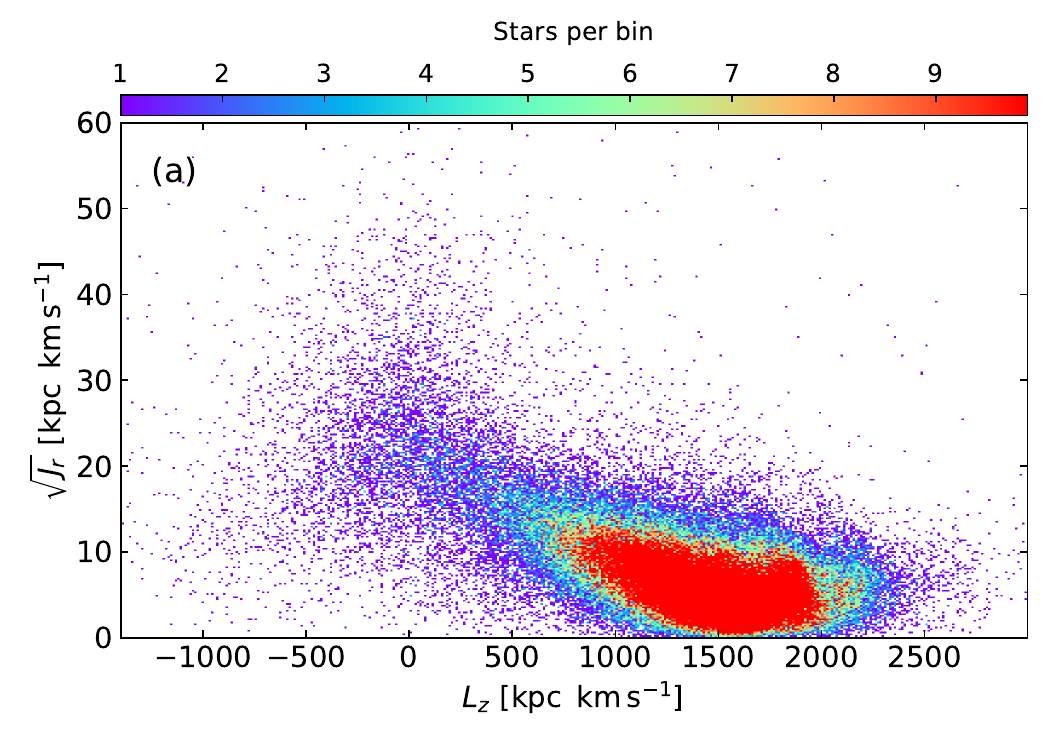}
   \includegraphics[viewport = 0  0 520 350,clip]{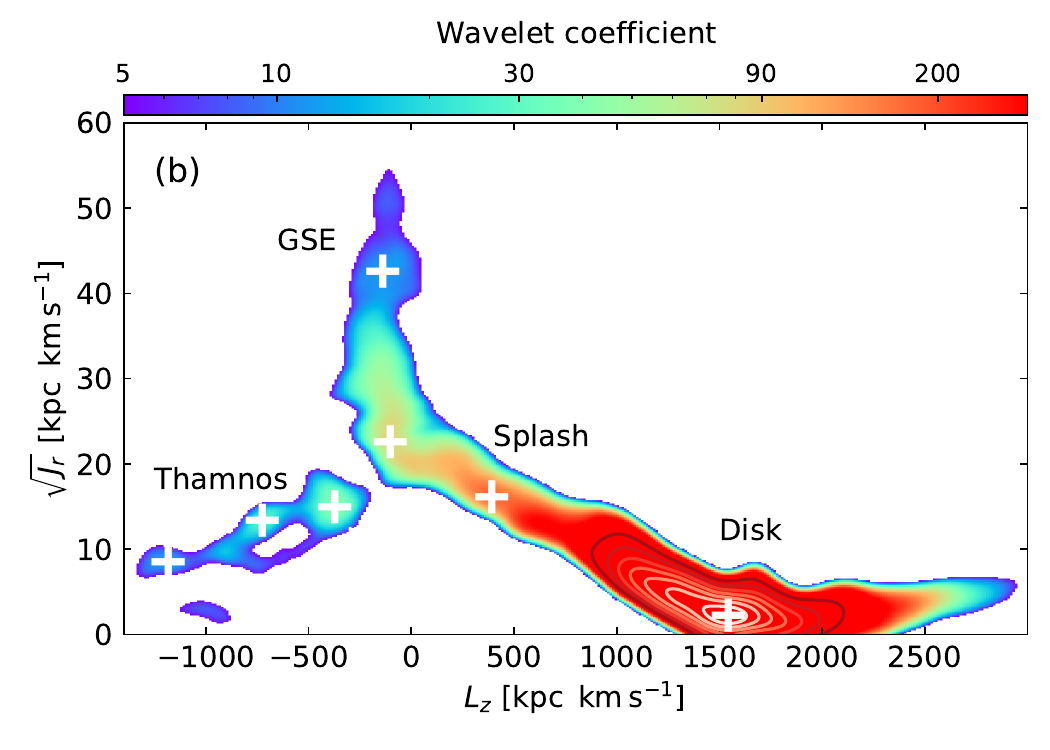}
   }
   \resizebox{\hsize}{!}{
   \includegraphics[viewport = 0  0 520 350,clip]{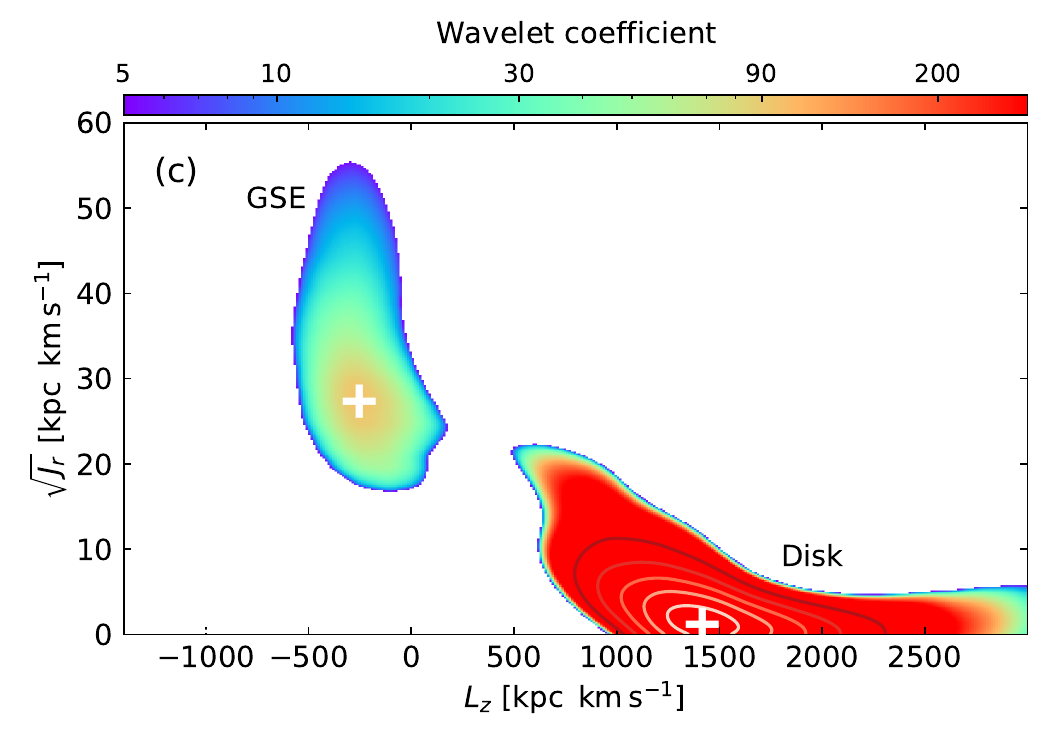}
   \includegraphics[viewport = 0  0 520 350,clip]{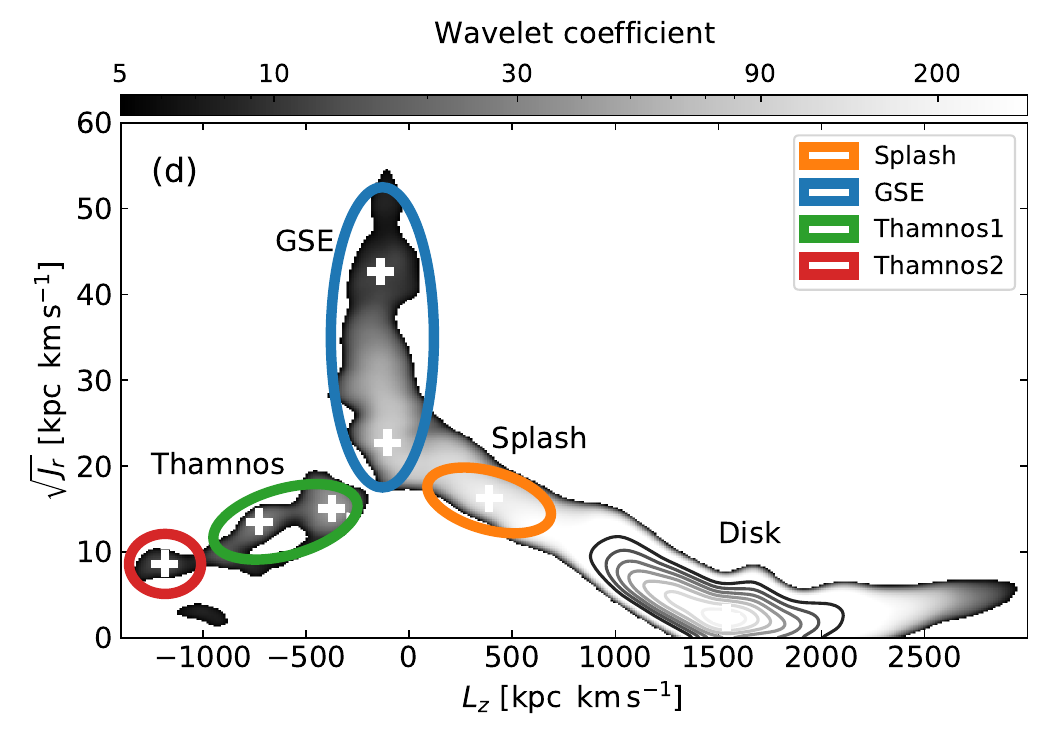}
   }
   \caption{
   Panel (a): Density map in $L_z - \sqrt{J_r}$ space of 124\,618 stars that were selected from GALAH DR4. The bin size is $11.7\times0.2$ kpc $\kms$.
   Panel (b): The wavelet transform map of 500 Monte-Carlo-generated samples in the $L_z - \sqrt{J_r}$ space for scale $J=5$. Centres of the detected structures are shown with white crosses. The structures are the disk, Splash, GSE, and Thamnos.
   Panel (c): The wavelet transform map of 500 Monte-Carlo-generated samples in the $L_z - \sqrt{J_r}$ space for scale $J=6$. Centres of the detected structures are shown with white crosses. GSE and the disk are the two main building blocks. 
   Panel (d): The same as panel (b), but in a gray scale. The ellipses show areas in the $L_z - \sqrt{J_r}$ space where we select stars to further analyze the detected kinematic structures. 
   \label{_fig_wavs}
   }
\end{figure*}

Panel (a) of Figure \ref{_fig_wavs} shows the initial 2D distribution of all 124\,618 stars in the $L_z - \sqrt{J_r}$ space. The Galactic disk is a dominant part of our sample and is seen at $\sqrt{J_r}$ < 20 kpc $\kms$ and $L_z$ > 1\,000 kpc $\kms$. The halo part accounts for fewer stars but is clearly visible in the sample. 

As the next step, we generate 500 Monte Carlo $L_z - \sqrt{J_r}$ maps of the initial 2D distribution, shown in panel (a) of Figure \ref{_fig_wavs} to which we apply the wavelet transform. This step is needed to check the significance of the detected overdensities. We calculated a covariance matrix between the parameters and found that most correlation coefficients are close to zero, indicating an insignificant dependence. Thus, we assume that a star's radial velocity, coordinates, proper motions, and distances follow a Gaussian distribution from which we draw values randomly and recalculate $L_z - \sqrt{J_r}$ 500 times. The mean values and widths of the Gaussian distributions are defined by the measured values and their corresponding uncertainties taken from the initial sample. Then, the wavelet transform is applied to every Monte Carlo-generated sample, and the central positions of all detected structures are over-plotted. Performing Monte Carlo simulations allows us to estimate the sizes and uncertainties of the detected kinematic groups. We found that 500 Monte Carlo-generated samples are enough to converge the results. 

\begin{table}
    \centering
    \caption{Detected kinematic structures identified in this work.}
    \begin{tabular}{rrrr}
    \hline
    \hline
    \noalign{\smallskip}
    Group     &$\sqrt{J_r}\pm\delta$ &${L_z}\pm\delta$ & Number of stars\\
              &[kpc $\kms$]          &[kpc $\kms$]     & \\
    \noalign{\smallskip}
    \hline
    \noalign{\smallskip}
    Splash   &            $16.0\pm4.5$   & $ 388\pm300$ & 3\,071\\
    GSE      &            $35.0\pm17.5$  & $-130\pm250$ & 1\,895\\
    Thamnos1 &            $13.5\pm4.0$   & $-600\pm350$ & 413   \\
    Thamnos2 &            $8.6\pm3.5$    & $-1184\pm175$& 29    \\
    \noalign{\smallskip}
    \hline
    \end{tabular}
    \label{_tab_groups}
\tablefoot{
Names of the structures are listed in Col.~1.
Cols.~2 and 3 give the centres of the ellipses drawn around the kinematic structures, from which stars were selected for further analysis (see Fig.~\ref{_fig_wavs}, panel~d).
The half-widths and half-heights of the ellipses ($\delta$) are given in Cols.~2 and 3.
Col.~4 provides the number of stars in each group.
}
\end{table}

The resulting wavelet transform maps for scales $J=5$ and $J=6$ are shown in panels (c) and (c) of Figure \ref{_fig_wavs} respectively. Both plots show sums of the wavelet transform maps of all 500 Monte Carlo-generated samples. With the help of an image processing package for Python called {\tt scikit-image}\footnote{\url{https://scikit-image.org}} \citep{_scikit} and its peak detection algorithm, we found seven overdensities for scale $J=5$ and two overdensities for scale $J=6$, which are marked with white crosses. Based on comparison of the results with the literature \citep[e.g.][]{_koppelman18, _feuillet20, _horta21} the disk, Splash, GSE (two crosses above $\sqrt{J_r}>20$ kpc $\kms$), Thamnos1, and Thamnos2 can be easily recognized based on their location on the $L_z - \sqrt{J_r}$ map. We found that GSE consists of two peaks on a scale of $J=5$, while Thamnos has three peaks that we distil into Thamnos1 and Thamnos2 groups based on previous studies. Next, we draw ellipses around the crosses as shown on panel (d) of Figure \ref{_fig_wavs} and select stars inside the ellipses to further analyze the detected kinematic structures. The ellipses were drawn based on a visual inspection of the structures, with a comparison of the resulting wavelet space to existing findings in the literature. Centers of the ellipses, widths and heights, and the number of stars inside each group we provide in Tab. \ref{_tab_groups}. The biggest ellipse belongs to GSE, for which we combined two peaks in scale $J=5$ as the structure is seen as a whole in scale $J=6$. We selected the most significant number of stars in the Splash population compared to the other groups. This reflects the dominance of the Solar neighborhood in the GALAH survey data.

\section{Contamination of the structures from the disk and halo}\label{_sec_contamination}

\begin{figure}
   \centering
   \resizebox{\hsize}{!}{
   \includegraphics[viewport = 0  0 520 350,clip]{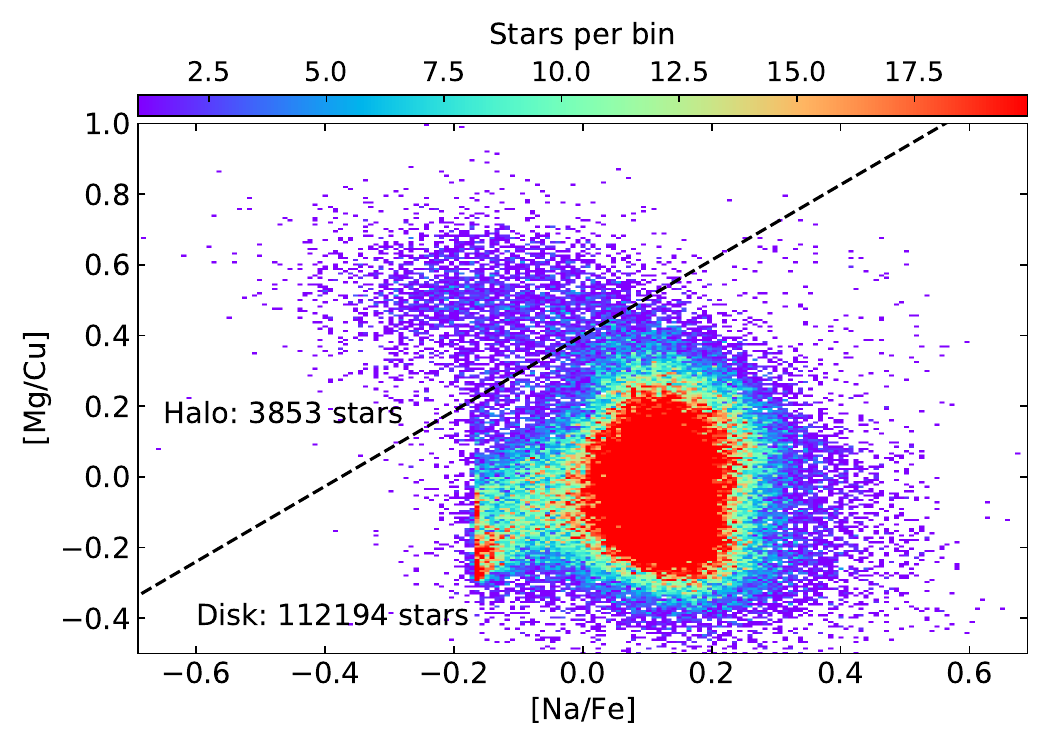}
   }
   \caption{\raggedright The binned distributions of 116 047 stars selected from the GALAH DR4 in the [Mg/Cu] – [Na/Fe] plane. The bin size is 0.01~×~0.01. The dashed line shows the division of stars into the disk (below the line) and halo (above the line) stars.
   \label{_fig_halo}
   }
\end{figure}

\begin{figure*}
   \centering
   \resizebox{\hsize}{!}{
   \includegraphics[viewport = 0  0 470 330,clip]{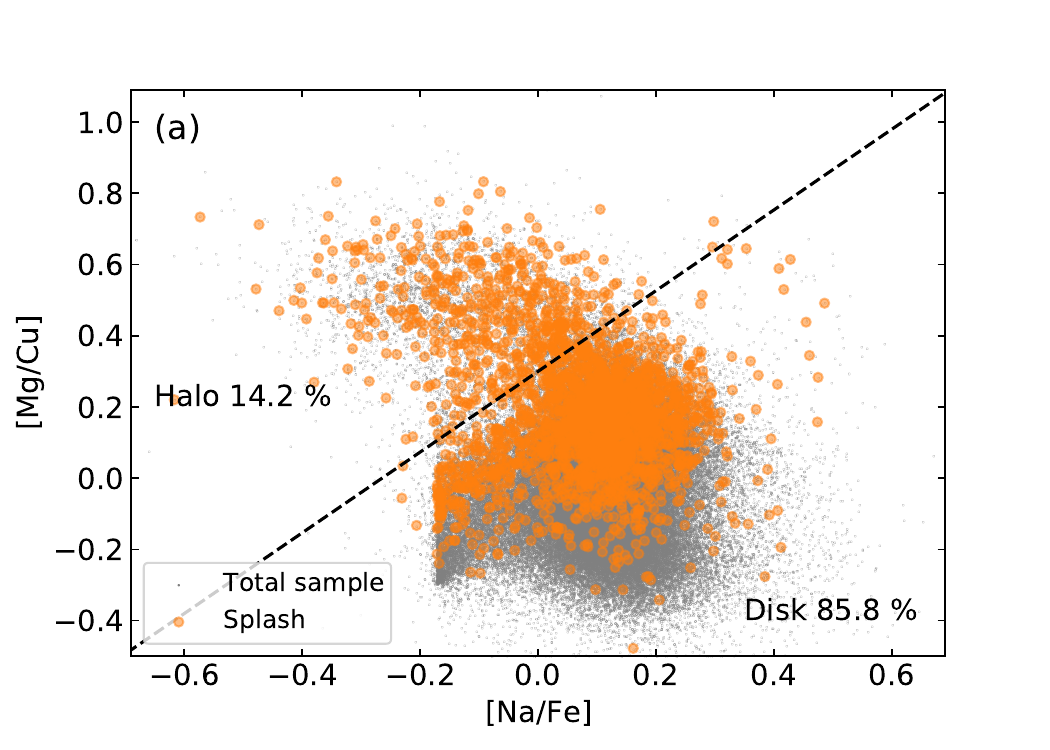}
   \includegraphics[viewport = 0  0 470 330,clip]{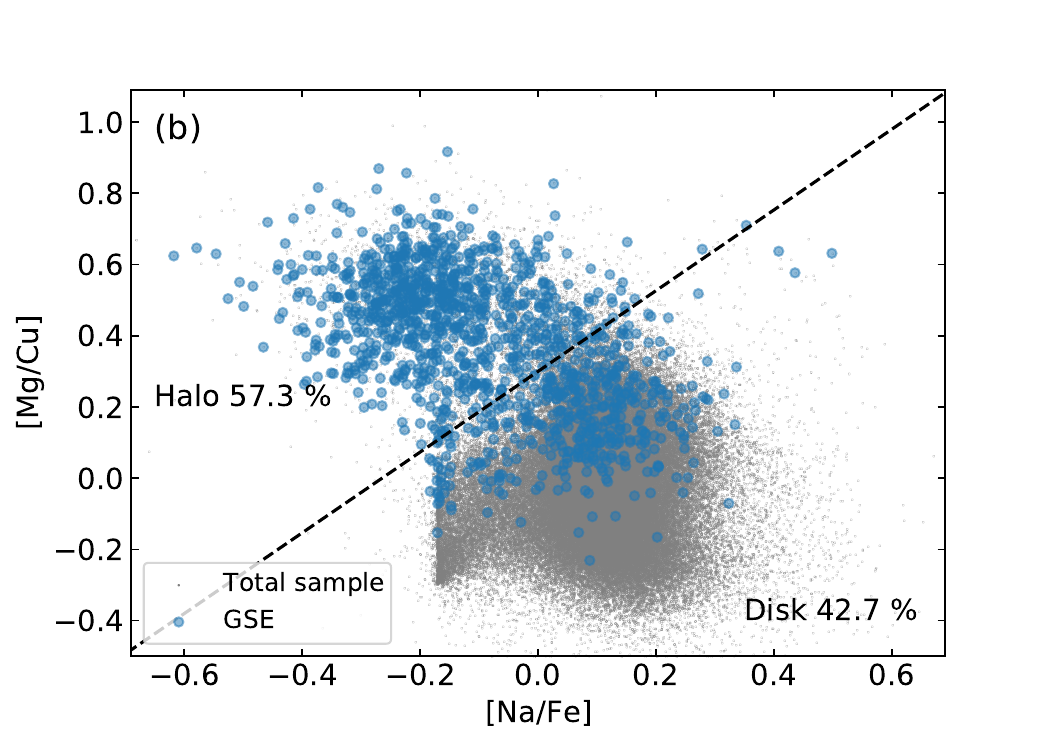}
   }
   \resizebox{\hsize}{!}{
   \includegraphics[viewport = 0  0 470 330,clip]{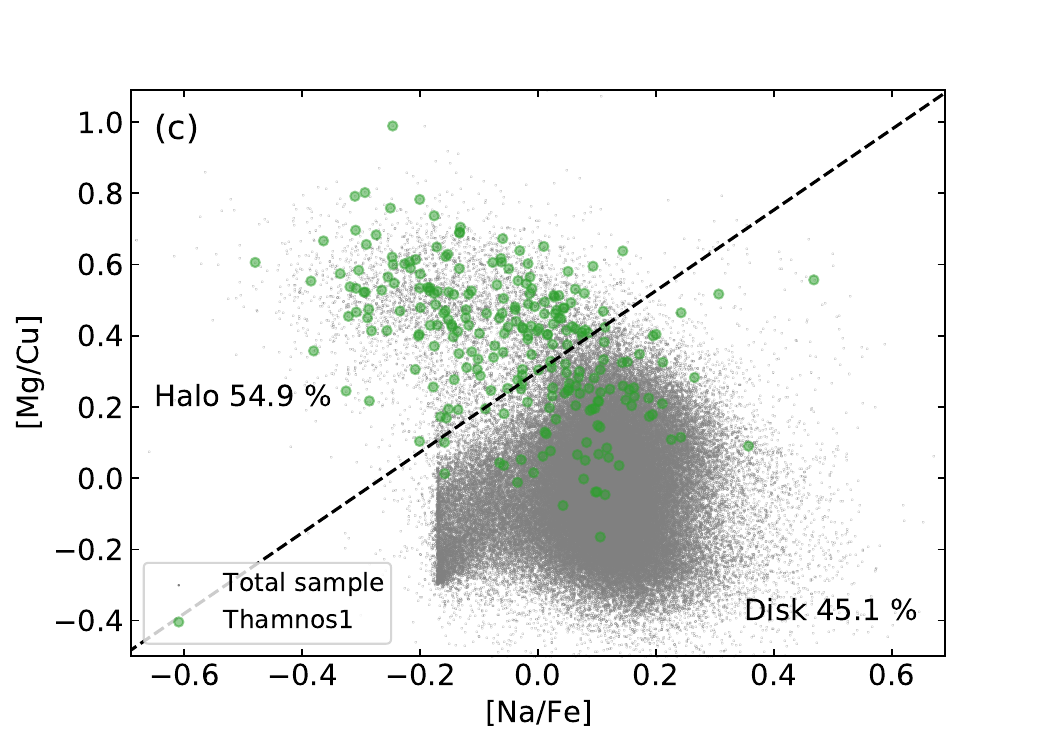}
   \includegraphics[viewport = 0  0 470 330,clip]{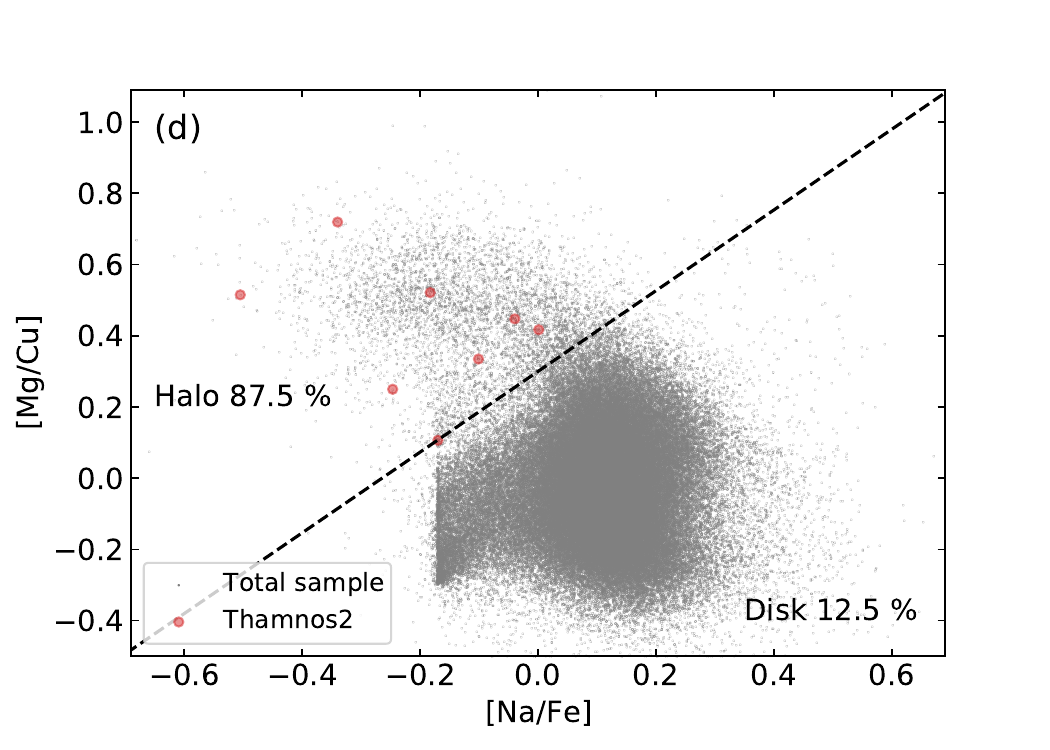}
   }
   \caption{A scatter plot of all 116\,047 stars selected from GALAH DR4 in the [Mg/Cu] -- [Na/Fe] plane is shown in gray. The markers of different colors show stars in the following kinematic structures: Splash (Panel (a), orange circles), GSE (Panel (b), blue circles), Thamnos1 (Panel (c), green circles), and Thamnos2 (Panel (d), red circles). The dashed line is the same as on panel (a) and divides the stars into the disk and halo sub-samples. Percentages of the halo and disk stars in the kinematic structures are provided in each plot. 
   \label{_fig_cu_fe}
   }
\end{figure*}

In recent studies, such as those conducted by \citet{_feuillet21}, \citet{_horta23}, and \citet{_feltzing23}, it was demonstrated that even though the kinematic and dynamical selection is efficient in identifying accreted halo populations, some "{\it in situ}" stars can still be included in the resulting sample. We check fractions of stars that belong to the chemically defined disk and halo in the total sample. To split stars into the disk and halo populations, we follow the approach used in \citet{_buder21} and study the distribution of stars in the [Mg/Cu] versus [Na/Fe]
space. [Mg/Cu] is a tracer of SNe II contributions from massive stars and SNIa of low-mass stars \citep{_kobayashi20}. Similarly to [Mg/Fe], [Na/Fe] allows tracking differences between the {\it in situ} and accreted stars \citep[e.g.][]{_nissen10}.

We also selected stars for which there are no identified problems with abundance determination of a chemical element X by setting the flag {\tt flag\_X\_fe} to zero. We applied this flag to each of the earlier-mentioned chemical elements to avoid inaccuracies in the analysis. The flags are given in the GALAH DR4 survey. Applying the flags to Mg, Cu, Na, and Fe abundances reduces the sample to 116\,047 stars. Figure \ref{_fig_halo} shows the fractions of stars belonging to the disk population (below the dashed line) and the halo (above the dashed line). The dashed line was visually determined after examining the density distribution. Figure \ref{_fig_cu_fe} shows the sample of 116\,047 GALAH stars in gray and the different colors and shapes show stars in the Splash, GSE, Thamnos1, and Thamnos2 structures. Some stars in the disk population are found to fall into the part of the plot around [Na/Fe]~$\simeq -0.17$, which is merely an artefact of the GALAH spectroscopic analysis pipeline. However, despite this, we do not exclude them from the study as most of these stars still belong to the disk population rather than the halo. Below, we discuss contamination fractions for each of the detected kinematic structures. 

\begin{itemize}
    \item Splash. By its definition, the Splash should be composed of stars with disk-like chemistry that were brought to halo-like orbits by the last major merger event \citep[see][]{_belokurov18}. According to this definition, we found that $\approx86\%$ of stars in the group we found can be categorized as chemically belonging to the disk, while the remaining $\approx14\%$ exhibit chemical characteristics associated with the halo.
    
    \item GSE. The GSE population originates from the last merger event and consists of accreted stars from the satellite Enceladus \citep{_helmi18}. Thus, GSE stars are expected to be chemically distinct from the disk stars. Panel (b) of Figure \ref{_fig_cu_fe} shows that $\approx57\%$ stars in the group we link to the GSE belong to the stellar halo, while the contamination from the disk is $\simeq 43\%$.
    
    \item Thamnos1. Thamnos1 is also believed to be a remnant of an accretion event and, thus, chemically differentiable from the disk \citep{_koppelman19}. In the plot on panel (c) of Figure \ref{_fig_cu_fe}, we see that the contamination from the disk is $\simeq 55\%$, and there is only $\simeq 45\%$ of halo stars.
    
    \item Thamnos2. Similarly to Thamnos1, Thamnos2 is expected to be clean from the disk stars. Panel (d) of Figure \ref{_fig_cu_fe} shows that $\simeq 88\%$ of the stars belong to the halo, and only one object is a disk star ($\simeq 12\%$).
\end{itemize}

Taking into account the definitions of the aforementioned kinematic structures and their contamination fractions, which are shown in Figure \ref{_fig_cu_fe}, we decided to exclude chemically-defined halo stars from the Splash and chemically-defined disk stars from the remaining four groups. 

\begin{figure*}
   \centering
   \resizebox{\hsize}{!}{
   \includegraphics[viewport = 0   5 470 330,clip]{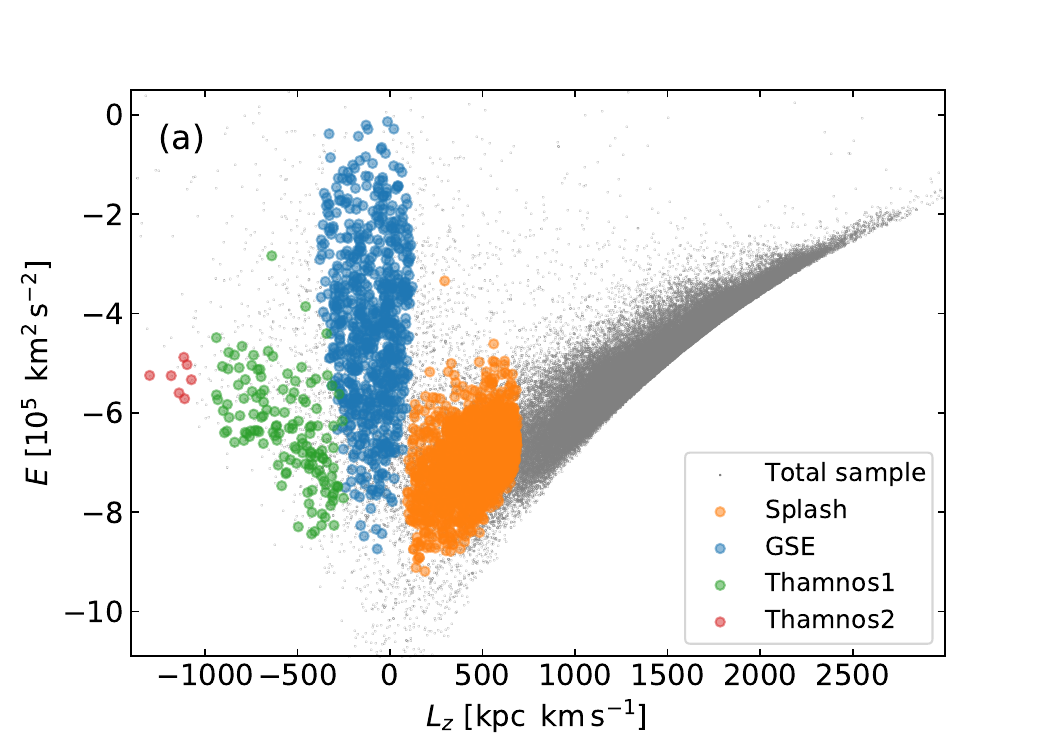}
   \includegraphics[viewport = 0  5 470 330,clip]{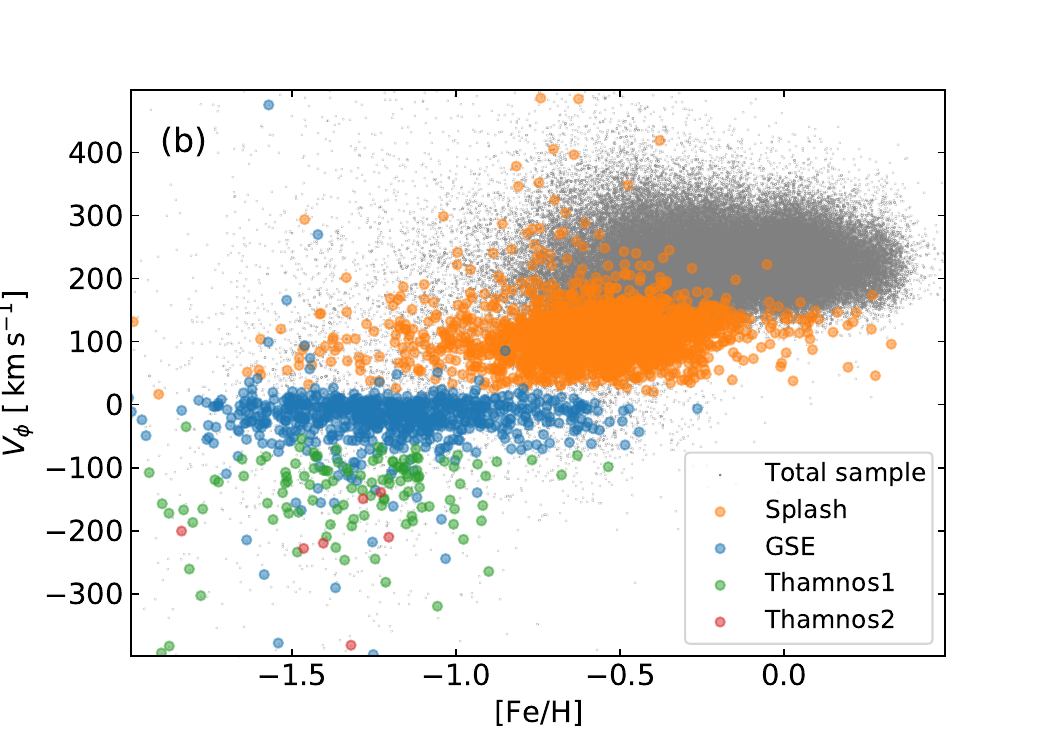}
   }
   \caption{Panel (a): The plot shows the orbital energy as a function of the angular momentum. Gray dots show a scatter plot of the total sample, and other colors and symbols correspond to the kinematic structures detected in the legend. Chemically-defined halo stars were excluded from the Splash, and chemically-defined disk stars were excluded from the rest of the groups. Panel (b): Similar to the plot on panel (a), but showing the rotational velocity as a function of metallicity. 
   \label{_fig_energy}
   }
\end{figure*}

Figure \ref{_fig_energy} shows the orbital energy distribution as a function of angular momentum and the rotational velocity versus metallicity for the entire sample and the kinematic structures detected. The results we observe agree with other studies that investigated similar distributions like \citet{_koppelman19} and \citet{_belokurov18}. The Splash is a connection between the Galactic disk stars and halo populations. The minor difference that we observe is that although stars in the GSE (Galactic substructure) population are close to $L_z\simeq0$ kpc $\kms$, they tend to follow slightly more retrograde orbits than prograde ones.

\section{Chemo-dynamical characterization of the groups}\label{_sec_characterization}

\begin{figure}
   \centering
   \resizebox{\hsize}{!}{
   \includegraphics[viewport = 0   5 470 330,clip]{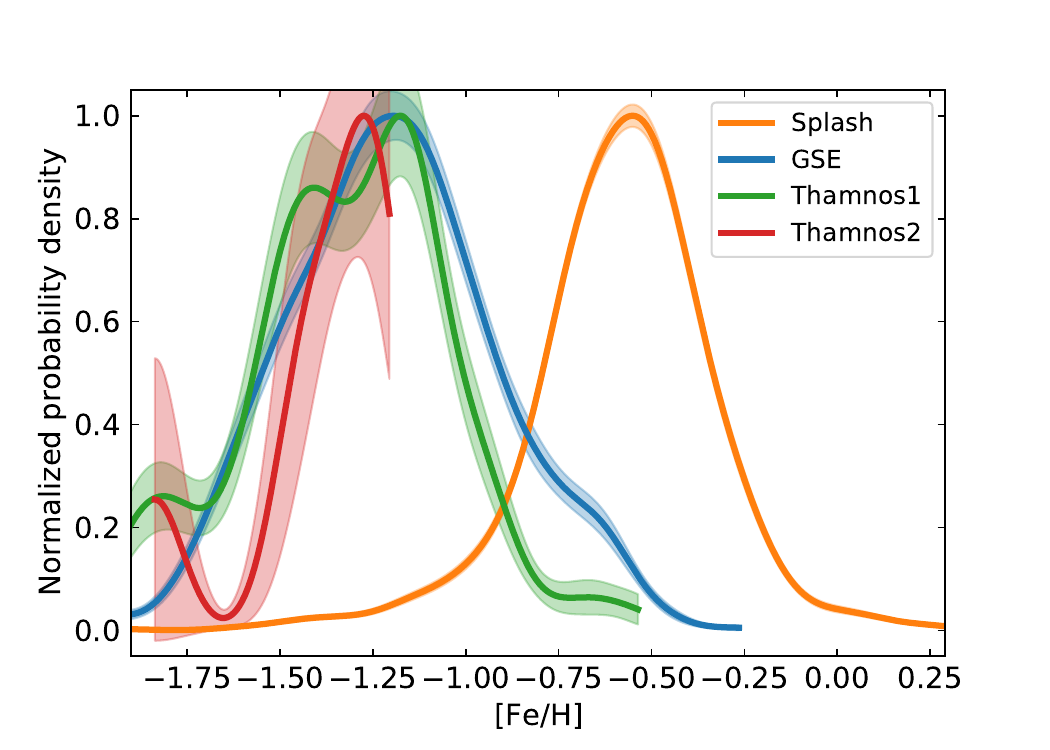}
   }
   \caption{Normalized probability density metallicity [Fe/H] with uncertainty bands for kinematic groups listed in the legend. The chemically-defined halo stars were excluded from the Splash, and chemically-defined disk stars were excluded from the rest of the groups. The plot is generated using kernel density estimation (KDE) with a bandwidth of 0.075 and Gaussian kernel. The shaded regions around each curve represent corresponding uncertainties, which are standard deviations of 100 bootstrap samples.
   \label{_fig_kde}
   }
\end{figure}

\begin{table}
    \centering
    \caption{KDE metallicity peak values for the detected kinematic structures.}
    \label{_tab_distr}
    \begin{tabular}{cccc}
    \hline
    \hline
    \noalign{\smallskip}
    Group     & [Fe/H]$_{\rm peak}$ & $\delta$[Fe/H] & Number of stars\\
    \noalign{\smallskip}
    \hline
    \noalign{\smallskip}
    Splash   &  $-0.55$ & 0.04 & 2\,200\\
    GSE      &  $-1.20$ & 0.07 & 748\\
    Thamnos1 &  $-1.17$ & 0.18 & 134\\
    Thamnos2 &  $-1.28$ & 0.79 & 7\\
    \noalign{\smallskip}
    \hline
    \end{tabular}
    \tablefoot{
The peak metallicity ([Fe/H]$_{\rm peak}$), corresponding uncertainties ($\delta$[Fe/H]), and the number of stars are listed for each kinematic group.
}
\end{table}

Figure \ref{_fig_kde} shows normalized metallicity distributions and corresponding uncertainties for the detected kinematic structures. These distributions were derived using kernel density estimation (KDE) algorithms from the {\tt scikit-learn}\footnote{\url{https://scikit-learn.org/}} package \citep{_scikit_learn}. Contamination of the halo and disk was removed for the Splash and the rest of the groups, respectively (see Sect.\ref{_sec_contamination}). Table \ref{_tab_distr} provides metallicity values, [Fe/H], where the normalized probability density has a maximum value and corresponding uncertainties, $\delta$. All groups have wide metallicity distributions and distributions of Galactocentric radii. Thamnos1 and Thamnos2 appear to have multiple-peaked distributions. This could be due to potential substructure within the groups or remaining contamination of the structures by field stars. GSE is the structure that covers the broadest range of Galactocentric radii and goes in the outer Galaxy (R $\gtrsim8$ kpc). The metallicity distribution of the GSE peaks at $\simeq\!\!-1.2$ and is in agreement with other studies like \citet{_naidy20}, who found a value of $\simeq\!\!-1.2$, and \citet{_feuillet21}, who found a mean value of $\simeq\!\!-1.15$. The peak of the metallicity distribution of Splash is $\simeq\!\!-0.6$, which is in agreement with, for example, \citet{_belokurov20} and \citet{_sanders21}. However, Thamnos1 and Thamnos2 have mean metallicities $\simeq\!\!-1.4$ according to \citet{_bellazzini23}, whereas, in our study, we found the peak of the metallicity distribution at $\simeq\!\!-1.2$ for Thamnos1 and $\simeq\!\!-1.3$ for Thamnos2, although these differences are still within the errors.  

\begin{figure*}
   \centering
   \resizebox{\hsize}{!}{
   \includegraphics[viewport = 10  10 1100 280,clip]{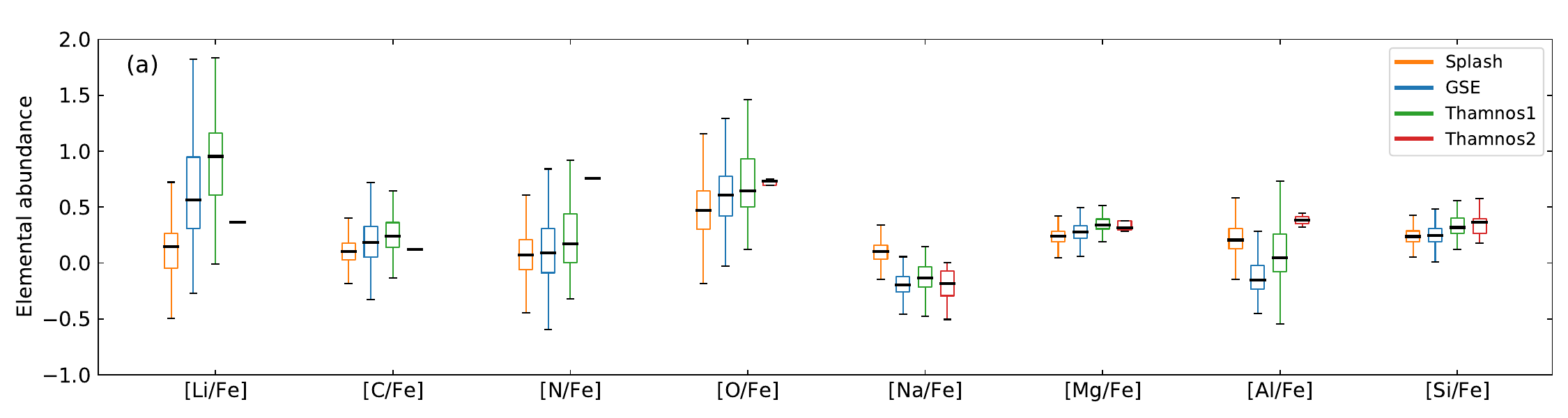}
   }
   \resizebox{\hsize}{!}{
   \includegraphics[viewport = 10  10  1100 280,clip]{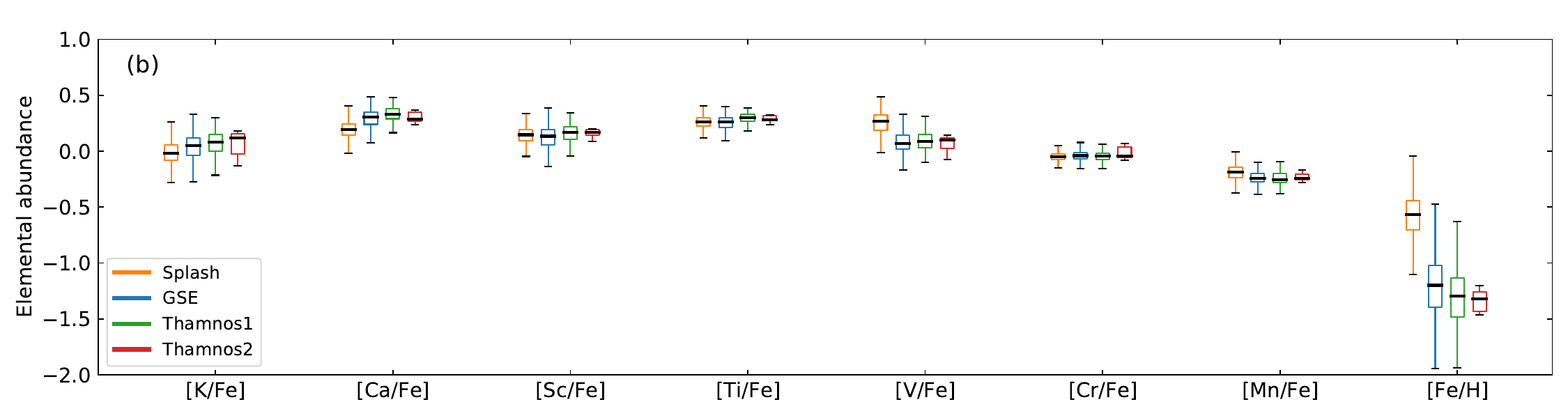}
   }
   \resizebox{\hsize}{!}{
   \includegraphics[viewport = 10  10  1100 280,clip]{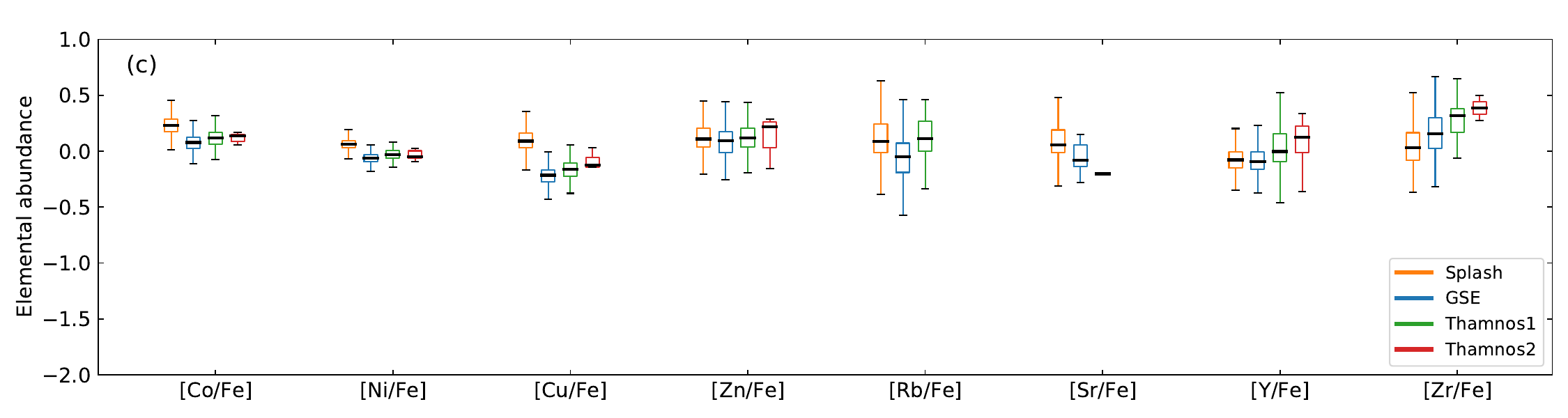}
   }
   \resizebox{\hsize}{!}{
   \includegraphics[viewport = 10  10  1100 280,clip]{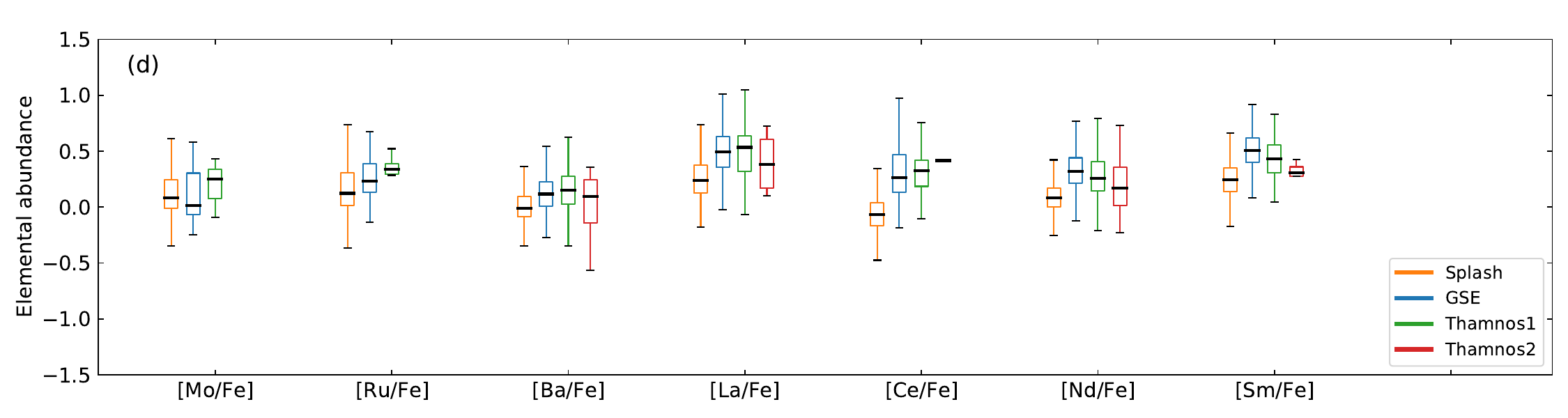}
   }
   \caption{Panels (a) -- (d) show individual distributions of elemental abundances for the Splash (orange box), GSE (blue box), Thamnos1 (green box), and Thamnos2 (red box) are depicted using box plots. The analysis covers 32 chemical elements, X, related to iron, [X/Fe], and iron abundance, which is related to hydrogen, [Fe/H]. Each box plot adheres to standard conventions, with the first and third quartiles defining the box body, "whiskers" extending to minimum and maximum values, and the median shown as a black line inside each box.
   \label{_fig_box}
   }
\end{figure*}

\begin{figure}
   \centering
   \resizebox{\hsize}{!}{
   \includegraphics[viewport = 0  0 430 280,clip]{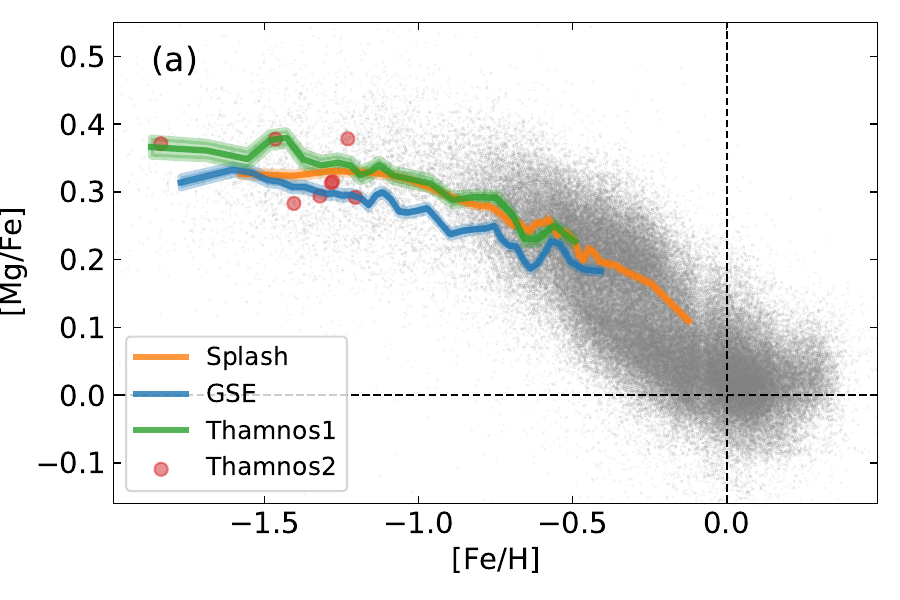}
   }
   \resizebox{\hsize}{!}{
   \includegraphics[viewport = 0  0 430 280,clip]{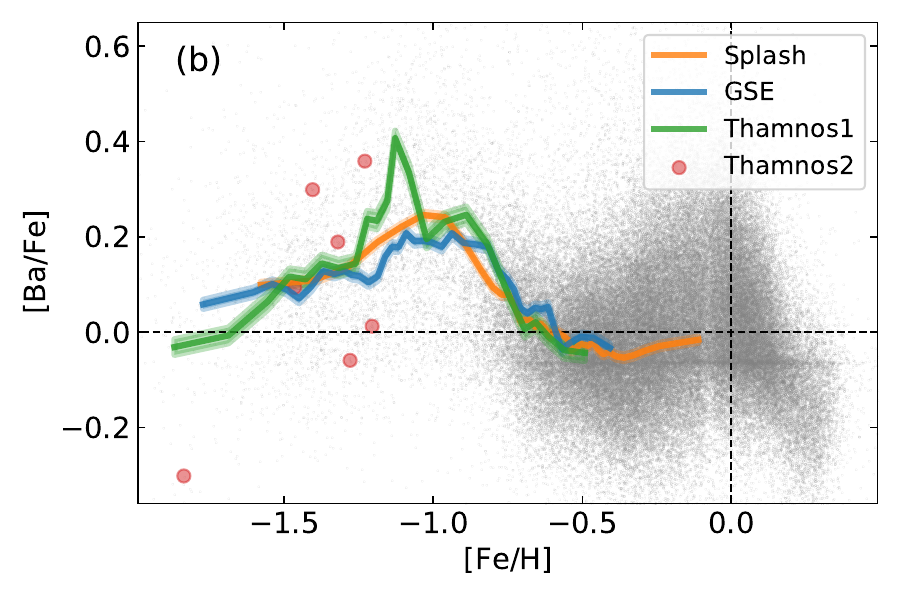}
   }
   \resizebox{\hsize}{!}{
   \includegraphics[viewport = 0  0 430 280,clip]{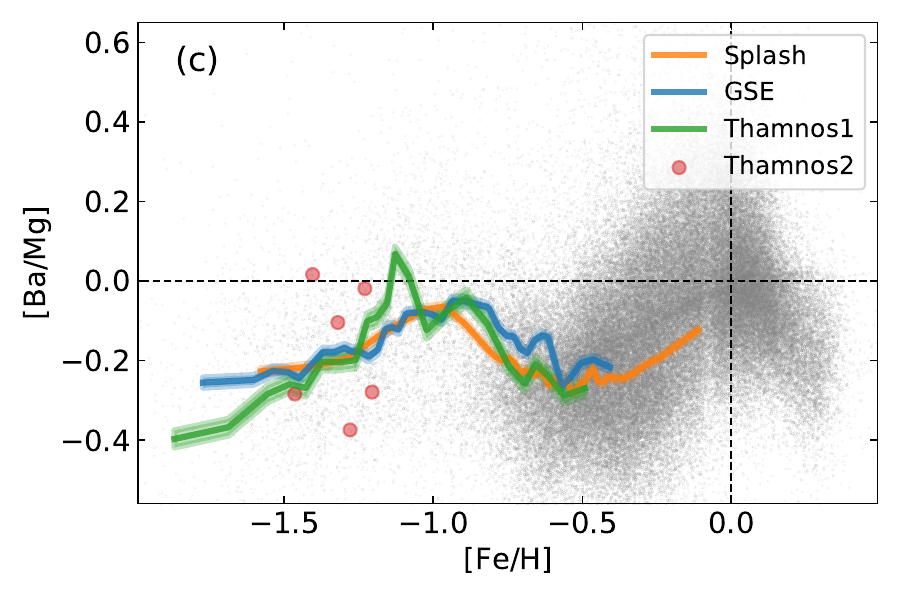}
   }
   \caption{Distribution of stars selected from GALAH DR4 in the following elemental abundance planes: (a) [Mg/Fe] -- [Fe/H], (b) [Ba/Fe] -- [Fe/H], (c) [Ba/Mg] -- [Fe/H]. The gray dots represent the total sample, while the colored lines show running means with shaded error regions for stars from kinematic structures as labeled in the legend. Stars in Thamnos2 are shown as dots due to the low number of stars in the group. The dashed lines indicate where the x- and y-axis are equal to zero.
   \label{_fig_chem}
   }
\end{figure}

Figure \ref{_fig_box} shows individual distributions of elemental abundances of all five kinematic structures, which are presented as box plots for 32 chemical elements. We included only stars for which individual elemental abundance flags, {\tt flag\_X\_fe} = 0, to get the most precise data for the box plots. Note that Mg, Fe, Cu, and Na abundances were used to identify populations, which means that we cannot draw purely independent conclusions based on these chemical elements. 

Figure \ref{_fig_chem} shows a more detailed view of several elemental abundance spaces, such as [Mg/Fe], [Ba/Fe], and [Ba/Mg] as a function of metallicity. All of these chemical spaces provide details about the star formation history of the detected kinematic groups, as all of these chemical elements have different origins. The running means of the detected structures are shown as colored lines, and the corresponding errors are shown as shaded regions. These shaded regions represent the standard error of the mean within each running window, calculated from the individual abundance uncertainties of the stars in that window. The total sample is shown in the background as gray dots for comparison. The window size for the running means was estimated based on the number of stars in each kinematic group and ranges from 5 to 50. The step size was adopted as half the window size. Since Thamnos2 has very few stars for the running means we show them as a scatter plot.

The uneven running mean of Thamnos1 results from several factors. Firstly, Thamnos1 has fewer identified stars compared to GSE and Splash. Secondly, the spike in Thamnos1 in panels (b) and (c) of Fig. \ref{_fig_chem} is likely due to contamination in the sample. This further emphasized the importance of combining both dynamical and chemical parameters to obtain the purest sample possible.

Below, we discuss elemental abundance patterns of each of the detected kinematic structures based on Figures \ref{_fig_box} and \ref{_fig_chem}.

\begin{itemize}
    \item The abundance of $\alpha$-elements such as C, O, Ca, Mg, Al, Si, and Ti are slightly higher in the halo structures. This is especially noticeable for Thamnos2, whose oxygen abundance is higher than that of the rest of the groups. This result is in agreement with \citet{_horta23}, who showed that Thamnos shows a higher [$\alpha$/Fe] ratio than the other halo substructures. 

    \item Iron-peak elements, including Co, Cr, Fe, Mn, Ni, Sc, and V, show different behaviours. Co, Fe, Ni, Sc, and V are more abundant in Splash than in the halo structures. The distributions of Cr and Mn are monotonic and do not vary between the groups. \citet{_sanders21} and later \citet{_nissen24} studied iron-peak abundances in the accreted halo populations and found low abundances of [Ni/Fe]. We observe a minor under-abundance of [Ni/Fe] in the accreted populations compared to Splash.

    \item Typical s-process (slow neutron-capture process) elements such as Ba, Ce, La, Mo, Nd, Rb, Sr, Y, and Zr are abundant in the accreted populations. The exception is Rb, whose abundance is lower for GSE than the rest of the structures, and Sr, for which very few stars in the structures have measurements. Thus, there are insufficient statistics for a proper analysis. However, \citet{_matsuno21} found a contribution of r-process elements dominates over s-process. This is in agreement with our results for other s-process elements in GSE like Ba (see Fig. \ref{_fig_chem}). The abundance of Ce, an s-process element, is much higher in GSE than in Splash, which contradicts the idea of low s-process abundances in GSE. \citet{_contursi23} found that Thamnos and GSE have similar Ce abundance, while our box plots show that GSE and Thamnos2 have a slightly lower median than Thamnos1. 

    \item GALAH DR4 provides abundances of the following r-process (rapid neutron-capture process) elements: Eu, Nd, and Sm. Eu is not used here due to its current unreliability in DR4. GSE tends to have a higher abundance of r-process elements than other structures. This result is also in agreement with previous findings on GSE \citep{_matsuno21}.

    \item The abundance of Cu, Ru, and Zn, elements produced by explosive and non-explosive nucleosynthesis, show different patterns. Copper is more abundant in Splash than the halo groups and is likely because we chemically defined disk and halo groups using Cu abundance (see Sect.\ref{_sec_contamination}). The abundance of Ruthenium is higher for GSE than Splash and Thamnos1. Zn is slightly more abundant in Thamnos1 and Thamnos2. 
    
    \item Abundances of K and Na, elements primarily produced through nucleosynthesis in stars. Both elements are slightly more abundant in the halo structures, although we used Na abundance to divide samples into the chemically defined disk and halo.

    \item Halo structures are more abundant in Li compared to Splash. The abundance of lithium in GSE was studied, for example, by \citet{_molaro20} and \citet{_simpson21}. It was shown that The A(Li) abundances versus [Fe/H] in the GSE population behave similarly to {\it in situ} stars, especially at low metallicities.

    \item The [Mg/Fe]--[Fe/H] space (see panel (a) in Fig.~\ref{_fig_chem}) shows a negative slope for all groups, reflecting the overall decline of $\alpha$-element abundances with increasing metallicity. The Splash occupies the thick disk region of the distribution, while GSE exhibits lower [Mg/Fe] values than both the Splash and Thamnos1. Thamnos1 follows a trend very similar to the Splash, indicating a later $\alpha$-knee and more efficient chemical enrichment. We note that possible contamination in the Thamnos1 selection may blur this trend. This behaviour is consistent with Thamnos1 having experienced more rapid star formation than GSE and is in agreement with, for example, \citet{_horta21}, who explored the chemical properties of kinematic substructures using the APOGEE survey.

    \item Thamnos1 and GSE show similar behavior in the [Ba/Fe]–[Fe/H] and [Ba/Mg]–[Fe/H] spaces (see panels (b), and (c) in Fig.~\ref{_fig_chem}), with both showing increasing trends with metallicity. The slope is steeper for Thamnos1 than for GSE, consistent with more rapid star formation and a stronger delayed contribution from AGB stars. The Splash exhibits a shallower slope, more similar to GSE, indicating that Thamnos1 may have undergone a chemical evolution different from both the Milky Way and GSE. We note that contamination and small number statistics may affect the precise slopes, but the overall trend is consistent with faster enrichment for Thamnos1.
\end{itemize}

Overall, accreted halo populations are chemically different from the disk stars. We also observed differences in elemental abundances among the dynamical populations, confirming that these groups have an extragalactic origin. Although the conclusion is, however, influenced by our chemical grouping, performing chemical, dynamical, and kinematic selection together is essential to neatly selecting stars that belong to the accreted groups. 

\section{Chemo-dynamical tagging of the accreted structures with t-SNE}\label{_sec_tsne}

Given that both dynamics and chemistry are essential for selecting halo substructures, we also investigate a method of selecting the halo substructures based solely on chemistry, in order to complement the dynamical selection using wavelets. To accomplish this, we use the algorithm t-distributed stochastic neighbor embedding (t-SNE), which implements dimensionality reduction based on non-linear manifold learning \citep{Hinton2002, vanderMaaten2008}. \citet{_kos18} showed the efficiency of this algorithm for identifying ejected stars from star clusters using chemical tagging with GALAH data, identifying two ejected stars several degrees from the main body of the Pleiades open cluster. A follow-up study by \citet{_youakim23} combined both chemistry and kinematics from GALAH data, using 20 parameters as input into t-SNE and found several tidally stripped stars from Omega Centauri, thereby constraining the cluster's initial mass and evolution time in the Galaxy. In another recent study conducted by \citet{_ortigoza23}, the t-SNE algorithm was used to investigate chemical structures in a sample of 1\,742 red giants with rotational velocity $V_{\phi} < 100 \kms $. The sample was selected from the APOGEE DR17 \citep{_abdurro22} and {\it Gaia} surveys. The input parameters of the t-SNE analysis included elemental abundances of 10 chemical elements. As a result of their study, they identified the GSE and Splash populations, as well as a few new structures. Taken together, these results demonstrate that the t-SNE algorithm is a helpful tool for grouping stars based on their chemical properties.

As input for the t-SNE algorithm in this study, we included measured abundances of 15 chemical elements: [Fe/H], [O/Fe], [Na/Fe], [Mg/Fe], [Al/Fe], [Si/Fe], [K/Fe], [Ca/Fe], [Sc/Fe], [Cr/Fe], [Mn/Fe], [Y/Fe], [Ba/Fe], and [Cu/Fe]. After application of the quality flags for each of these elements ({\tt flag\_X\_fe} = 0), as well as removal of any star missing data for any of these abundances (we only included measurements, not upper limits), the sample was reduced to 107 769 stars.

In order to mitigate the different ranges of these abundances, we standardized each input distribution such that they were transformed to have a median of 0 and a standard deviation of 1. In practice, we accomplished this by subtracting each abundance value from the median and dividing by the standard deviation. The abundance distributions are close enough to being Gaussian that this simple standardization was sufficient. We also applied weights in order to increase the contribution from certain abundances. Weights for abundances were selected to increase the contribution of elements that are known from the literature to be important for selecting halo substructures, such as [Fe/H] and [Cu/Fe], and to account for the observational uncertainties, where the weights of abundances with large observational uncertainties are reduced. A thorough investigation of the weights for different abundances in GALAH was conducted in \citet{_kos18}, and we refer the reader there for a more detailed description. In this work, we adopt the weights used in that paper, and all of the weights for the input parameters are summarized in Table \ref{_tab_weights}. We also checked the analysis using uniform weights, and found that the structures were still localized in the latent space, but with a slightly higher contamination.

\begin{figure*}
   \centering
   \includegraphics[width=\textwidth]{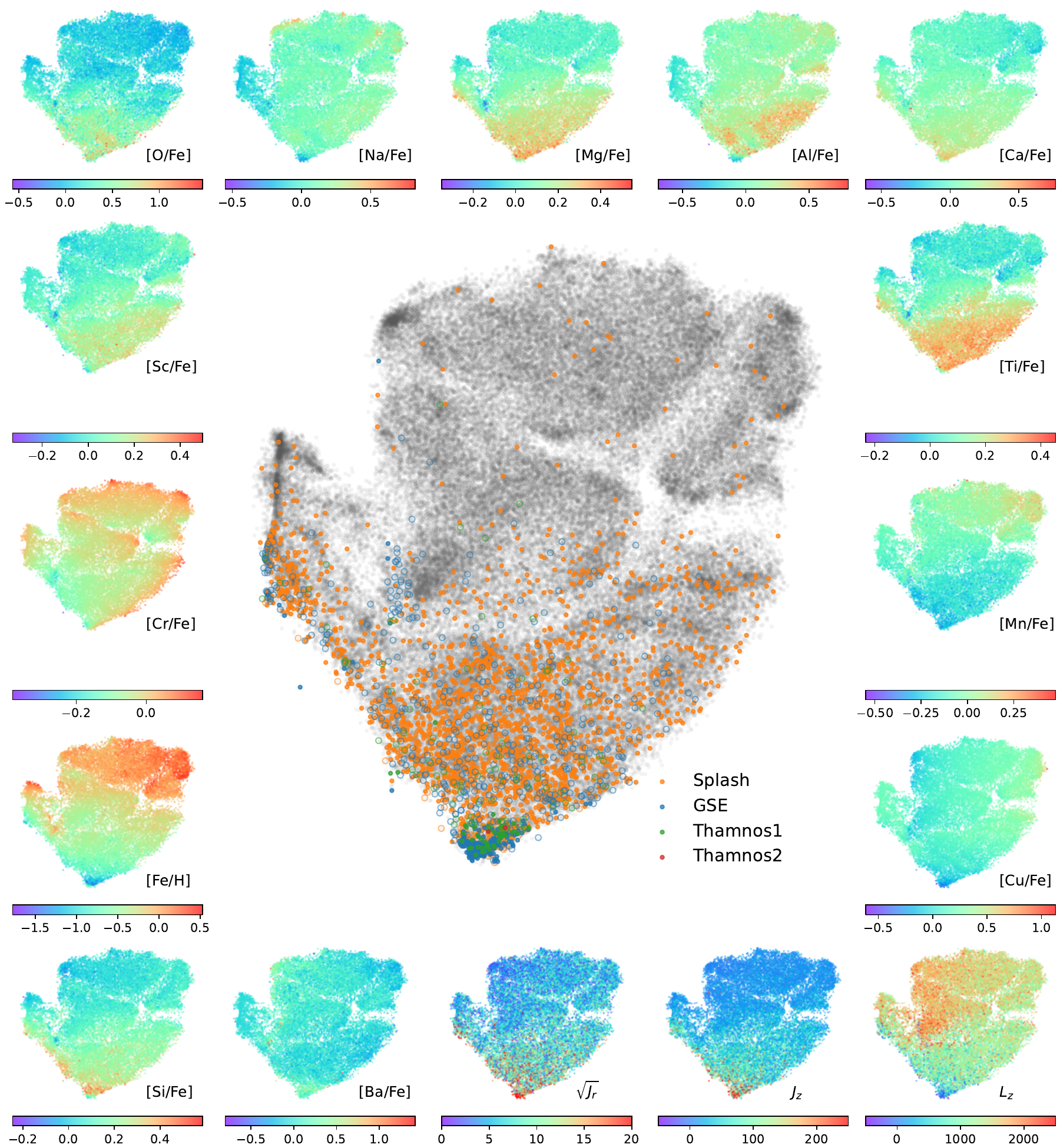}
   \caption{t-SNE latent space projection of all stars in the GALAH sample, shown in gray. Halo stars from GSE and Thamnos1 and 2 are shown in blue, green and red, respectively, while disk stars from the Splash are shown in orange. Filled points show stars identified as belonging to the halo based on the cuts applied in Figure \ref{_fig_cu_fe}, while open circles are stars selected based on their dynamics but did not pass this chemical selection. The surrounding plots show the same latent space projection, colored by various parameters.}
   \label{_fig_tsne}
\end{figure*}

The middle panel of Figure \ref{_fig_tsne} shows the two-dimensional latent space projection resulting from the dimensionality reduction. There are two input hyper-parameters that we used to initiate the t-SNE, n\_jobs, which defines how many CPU cores are dedicated for the computation and for which we used a value of 4, and perplexity, which is a measure of the expected size of the groups that cluster in the latent space. Lower perplexity values allow to highlight smaller groups, while higher values help to detect large-scale structure. After probing a range of values, we chose a perplexity of 1\,000 as input into the algorithm. Smaller values tended to separate the labeled GSE stars into several smaller substructures, and values much larger than this resulted in a single uniform cluster of all of the stars in the sample. The filled colored points show the selected stars using the wavelet analysis, limited to candidates that are also consistent with being halo stars for GSE and Thamnos, and limited to disk stars for the Splash (open circles are the disk stars that were removed according to the selection shown in Figure \ref{_fig_cu_fe}). Stars selected as GSE stars in the wavelet analysis (blue points) cluster very tightly in the t-SNE projection, demonstrating that these can be identified from field stars and disk stars based solely on their elemental abundance patterns. However, the GSE stars overlap substantially with the Thamnos1 and Thamnos2 stars (green and red points), suggesting that they are not chemically distinct enough, or that the GALAH abundances are not precise enough, to differentiate these structures through chemical tagging alone. The blue and green open circles which scatter up the bottom half of the t-SNE map indicate that the cut made to select halo stars in Figure \ref{_fig_cu_fe} is very effective at removing contaminants from a purely kinematic selection.

Surrounding the main t-SNE latent space projection in Figure \ref{_fig_tsne}, there are smaller representations of the same latent space, but color coded by different parameters. This is very insightful to see which abundances are the most important in the identification of these halo substructures. We note that we have not plotted the maps for [K/Fe] and [Y/Fe], since both of these were very uniform and contained little information, with [Y/Fe] looking exactly the same as [Ba/Fe], since these elements share similar chemical evolutionary pathways. Instead, we include the dynamical parameters $\sqrt{J_r}$, $J_z$ and $L_z$, which were not included as inputs into the t-SNE. Notable elements include [Fe/H], with a clear metallicity gradient showing in the latent space, [Cu/Fe], as well as alpha elements [Mg/Fe], [Ca/Fe], [Ti/Fe], and the light elements [Na/Fe] and [Al/Fe], all of which previously have been shown to be important tracers in Galactic chemical evolution. Finally, all of the dynamical parameters $\sqrt{J_r}$, $J_z$ and $L_z$ show a clearly visible feature at the location of GSE and Thamnos in the latent space, confirming that there is a distinct overlap in the chemical and kinematic signatures of these structures.

We note that there are several other obvious overdensities located in the disk stars shown in gray in Figure~\ref{_fig_tsne}. We investigated these and found that most of them are kinematic moving groups in the disk, like Hercules, Sirius, Coma Berenices, Pleiades, Hyades and others, that have previously been discussed in the literature \citep[ex.:][]{_antoja12, _ramos18, _kushniruk17, _kushniruk19} or some remnants of globular cluster stars in the outskirts of the cluster that were not removed by the selection criteria. A proper analysis of these structures is beyond the scope of this paper, but demonstrates the efficiency of t-SNE at identifying chemical structures in the Galaxy.

\begin{table}
\centering
\caption{List of all of the weights used for the standardized input t-SNE parameters.}
\begin{tabular}{cccccc}
    \hline
    \hline
    \noalign{\smallskip}
Param & Weight & Param & Weight & Param               & Weight \\
    \noalign{\smallskip}
    \hline
    \noalign{\smallskip}
O     & 1.0    & Ti    & 2.0    & Ca               & 0.5    \\ 
Na    & 1.0    & Cr    & 2.0    & Y               & 0.25    \\
Mg    & 0.50   & Mn    & 1.0    & Sc               & 1.0    \\
Al    & 1.0    & Fe    & 2.0    & Ba               & 0.25    \\
K     & 0.25   & Cu    & 2.0    & Si               & 1.0    \\
    \noalign{\smallskip}
    \hline
\end{tabular}
\label{_tab_weights}
\end{table}

\section{Summary}\label{_sec_summary}

In this work, we analyzed a sample of 124\,618 carefully selected red giants from the GALAH DR4 and {\it Gaia} surveys using wavelet transform and t-SNE algorithms. Our analysis revealed the following:

\begin{itemize}
    \item Using the wavelet transform technique, we were able to detect the main structures of a space defined by a square root of radial action and angular momentum, namely the Galactic disk, Splash, GSE, Thamnos1 and Thamnos2. 
    \item The strongest detection in the wavelet maps corresponds to the Galactic disk. This is due to a large number of disk stars (112\,194) in our sample. The t-SNE algorithm revealed dozens of disk structures, among which we could recognize well-known moving groups in the Galactic disk.  
    \item The Splash population on the wavelet transform map plays the role of a bridge that links the metal-rich part of the Galactic disk and GSE. Because of proximity to accreted populations in action space, these {\it in situ} stars on halo-like orbits contaminate them if selected purely based on kinematics. On the other hand, the Splash population is also contaminated by the halo stars. We observe only one peak that we link to Splash. 
    \item Our findings indicate that GSE can be divided into two substructures: the blob of accreted stars at $J_r > 40$ kpc $\kms$ and the blob at lower values of the square root of the radial action, which is a mixture of {\it in situ} and accreted stars. We also observe that the GSE population is not purely a non-rotating component of the Galaxy. Wavelet transform shows that although close to $Lz\simeq0$ kpc $\kms$, many stars follow slightly retrograde orbits. 
    \item In Thamnos, we discovered three peaks in the action space. Two of these peaks are combined and referred to as the Thamnos1 group. A kinematic selection of stars in the Thamnos1 group indicates that a substantial number of them belong to the high-alpha disk. However, these stars can be removed from the sample if the elemental abundances of individual stars are studied. Thamnos2, on the other hand, is mainly composed of halo stars.
    \item The analysis of our sample with the t-SNE method using only chemical abundances as input successfully recovers the groups of stars identified with the wavelet transform. Chemical tagging could differentiate GSE, Thamnos1 and Thamnos2 from the bulk Milky Way population, but could not separate these halo structures from each other based on chemistry alone. This indicates the importance of kinematics for identifying and separating differentiating these structures. The t-SNE analayis also showed that the halo selection in [Mg/Cu] vs. [Na/Fe] space is very effective at removing contamination when selecting stars in GSE, Thamnos1 and Thamnos2. The splashed disk is located in immediate proximity to the accreted group of stars, indicating that the Splash contaminates the selection of accreted populations. 
\end{itemize}

The analysis of chemical properties of the stars in the kinematic groups is partly influenced by the way we defined the groups chemically based on Mg, Fe, Cu and Na elemental abundances, but still allows us to conclude the following:

\begin{itemize}
    \item Although a kinematic selection only allows the choice of stars in the accreted populations, an analysis of the chemical properties of these stars is needed for the 'purest' selection. The structures close to the $L_z$ and with $\sqrt{J_r} < 25$ kpc $\kms$ have a higher contamination rate.  
    \item The peak values of metallicity distributions for each group are in agreement with the previous studies and place Splash at $-0.6$ dex, GSE at $-1.2$ dex, Thamnos1 at $-1.2$, Thamnos2 at $-1.3$ dex. Peaks of the Galactocentric radius distributions fall into the inner Galaxy, and the GSE population covers the broadest range of Galactic radii and extends at least up to 15 kpc from the Galactic center.
    \item Stars in the Splash exhibit chemical properties of the Galactic disk, among which is high metallicity above $-0.5$ dex and a relatively high abundance of iron-peak elements. 
    \item Chemical peculiarities of GSE include low-metallicity, high [Li/Fe] abundance and high abundance of an r-process element Sm. Surprisingly, we observe a relatively high abundance of some s-process elements, such as Ce in GSE, which does not agree with previous studies like \citet{_matsuno21}. 
    \item Thamnos has a slightly elevated [$\alpha$/Fe] ratio than the rest of the accreted groups. Among other chemical peculiarities of the Thamnos population are abundances of Li, Ru, and Zr, which are slightly elevated than those of the rest of the groups. 
\end{itemize}

In recent years, significant effort has been made to identify and study the chemo-dynamical properties of the inner halo structures. The GALAH and {\it Gaia} surveys play a crucial role in searching for and analyzing kinematic parameters and elemental abundances of stars within accreted stellar populations. With these surveys, the remnants of accreted populations can be easily found in the Solar vicinity. Great hope relies on future spectroscopic programs to cover a broader range of Galactocentric distances, reaching as far as $R > 20$ kpc to study the outer parts of the stellar halo. The high-precision spectroscopic data of future missions like 4MOST \citet{_dejong2019} and WEAVE \citet{_dalton18, _dalton20} will hopefully shed more light on the nature of these unique building blocks of our Galaxy. 

\section*{Data availability}\label{_availability}
The GALAH DR4 catalog is available at \url{https://www.galah-survey.org/dr4/overview/} and at the Strasbourg Astronomical Data Center (CDS) via \url{https://cds.unistra.fr}. We provide a supplementary table containing information for 116\,047 stars with GALAH DR4 and Gaia DR3 identifiers. The column ``halo'' indicates whether a star belongs to the stellar halo (True) or the disk (False) based on [Mg/Cu] vs. [Na/Fe] criteria. The columns ``Splash'', ``GSE'', ``Thamnos1'', and ``Thamnos2'' indicate whether a star belongs to the corresponding kinematic structure (True or False) based on the wavelet transform. The table is only available in electronic form and can be retrieved from the CDS via anonymous ftp to \url{cdsarc.u-strasbg.fr} (130.79.128.5) or via \url{http://cdsweb.u-strasbg.fr/cgi-bin/qcat?J/A+A/}.

\begin{acknowledgements}
We acknowledge the traditional owners of the land on which the AAT stands, the Gamilaraay people, and pay our respects to elders past and present. This work was partly supported by the Australian Research Council Centre of Excellence for All Sky Astrophysics in 3 Dimensions (ASTRO 3D) through project number CE170100013. IK, KY, and KL acknowledge funding from the European Research Council (ERC) under the European Union's Horizon 2020 research and innovation program (Grant Agreement No. 852977)". SB acknowledges support from the Australian Research Council under grant number DE240100150. JBH is funded by an ARC Laureate Fellowship. ZT acknowledges support from the Slovenian Research Agency (core funding No. P1-0188). DZ is funded by ARC Discovery Project DP220102254. SLM acknowledges the support of the Australian Research Council through Discovery Project grant DP220102254 and the UNSW Scientia Fellowship Program. 
\end{acknowledgements}


\bibliographystyle{aa}
\bibliography{references}

@ARTICLE{_buder22,
       author = {{Buder}, Sven and {Lind}, Karin and {Ness}, Melissa K. and {Feuillet}, Diane K. and {Horta}, Danny and {Monty}, Stephanie and {Buck}, Tobias and {Nordlander}, Thomas and {Bland-Hawthorn}, Joss and {Casey}, Andrew R. and {de Silva}, Gayandhi M. and {D'Orazi}, Valentina and {Freeman}, Ken C. and {Hayden}, Michael R. and {Kos}, Janez and {Martell}, Sarah L. and {Lewis}, Geraint F. and {Lin}, Jane and {Schlesinger}, Katharine J. and {Sharma}, Sanjib and {Simpson}, Jeffrey D. and {Stello}, Dennis and {Zucker}, Daniel B. and {Zwitter}, Toma{\v{z}} and {Ciuc{\u{a}}}, Ioana and {Horner}, Jonathan and {Kobayashi}, Chiaki and {Ting}, Yuan-Sen and {Wyse}, Rosemary F.~G. and {Wyse}, Galah Collaboration},
        title = "{The GALAH Survey: chemical tagging and chrono-chemodynamics of accreted halo stars with GALAH+ DR3 and Gaia eDR3}",
      journal = {\mnras},
         year = 2022,
        month = feb,
       volume = {510},
       number = {2},
        pages = {2407-2436},
          doi = {10.1093/mnras/stab3504},
archivePrefix = {arXiv},
       eprint = {2109.04059},
 primaryClass = {astro-ph.GA},
       adsurl = {https://ui.adsabs.harvard.edu/abs/2022MNRAS.510.2407B}
}

@ARTICLE{_gaiadr3,
       author = {{Gaia Collaboration} and {Vallenari}, A. and {Brown}, A.~G.~A. and {Prusti}, T. and {de Bruijne}, J.~H.~J. and {Arenou}, F. and {Babusiaux}, C. and {Biermann}, M. and {Creevey}, O.~L. and {Ducourant}, C. and {Evans}, D.~W. and {Eyer}, L. and {Guerra}, R. and {Hutton}, A. and {Jordi}, C. and {Klioner}, S.~A. and {Lammers}, U.~L. and {Lindegren}, L. and {Luri}, X. and {Mignard}, F. and {Panem}, C. and {Pourbaix}, D. and {Randich}, S. and {Sartoretti}, P. and {Soubiran}, C. and {Tanga}, P. and {Walton}, N.~A. and {Bailer-Jones}, C.~A.~L. and {Bastian}, U. and {Drimmel}, R. and {Jansen}, F. and {Katz}, D. and {Lattanzi}, M.~G. and {van Leeuwen}, F. and {Bakker}, J. and {Cacciari}, C. and {Casta{\~n}eda}, J. and {De Angeli}, F. and {Fabricius}, C. and {Fouesneau}, M. and {Fr{\'e}mat}, Y. and {Galluccio}, L. and {Guerrier}, A. and {Heiter}, U. and {Masana}, E. and {Messineo}, R. and {Mowlavi}, N. and {Nicolas}, C. and {Nienartowicz}, K. and {Pailler}, F. and {Panuzzo}, P. and {Riclet}, F. and {Roux}, W. and {Seabroke}, G.~M. and {Sordo}, R. and {Th{\'e}venin}, F. and {Gracia-Abril}, G. and {Portell}, J. and {Teyssier}, D. and {Altmann}, M. and {Andrae}, R. and {Audard}, M. and {Bellas-Velidis}, I. and {Benson}, K. and {Berthier}, J. and {Blomme}, R. and {Burgess}, P.~W. and {Busonero}, D. and {Busso}, G. and {C{\'a}novas}, H. and {Carry}, B. and {Cellino}, A. and {Cheek}, N. and {Clementini}, G. and {Damerdji}, Y. and {Davidson}, M. and {de Teodoro}, P. and {Nu{\~n}ez Campos}, M. and {Delchambre}, L. and {Dell'Oro}, A. and {Esquej}, P. and {Fern{\'a}ndez-Hern{\'a}ndez}, J. and {Fraile}, E. and {Garabato}, D. and {Garc{\'\i}a-Lario}, P. and {Gosset}, E. and {Haigron}, R. and {Halbwachs}, J. -L. and {Hambly}, N.~C. and {Harrison}, D.~L. and {Hern{\'a}ndez}, J. and {Hestroffer}, D. and {Hodgkin}, S.~T. and {Holl}, B. and {Jan{\ss}en}, K. and {Jevardat de Fombelle}, G. and {Jordan}, S. and {Krone-Martins}, A. and {Lanzafame}, A.~C. and {L{\"o}ffler}, W. and {Marchal}, O. and {Marrese}, P.~M. and {Moitinho}, A. and {Muinonen}, K. and {Osborne}, P. and {Pancino}, E. and {Pauwels}, T. and {Recio-Blanco}, A. and {Reyl{\'e}}, C. and {Riello}, M. and {Rimoldini}, L. and {Roegiers}, T. and {Rybizki}, J. and {Sarro}, L.~M. and {Siopis}, C. and {Smith}, M. and {Sozzetti}, A. and {Utrilla}, E. and {van Leeuwen}, M. and {Abbas}, U. and {{\'A}brah{\'a}m}, P. and {Abreu Aramburu}, A. and {Aerts}, C. and {Aguado}, J.~J. and {Ajaj}, M. and {Aldea-Montero}, F. and {Altavilla}, G. and {{\'A}lvarez}, M.~A. and {Alves}, J. and {Anders}, F. and {Anderson}, R.~I. and {Anglada Varela}, E. and {Antoja}, T. and {Baines}, D. and {Baker}, S.~G. and {Balaguer-N{\'u}{\~n}ez}, L. and {Balbinot}, E. and {Balog}, Z. and {Barache}, C. and {Barbato}, D. and {Barros}, M. and {Barstow}, M.~A. and {Bartolom{\'e}}, S. and {Bassilana}, J. -L. and {Bauchet}, N. and {Becciani}, U. and {Bellazzini}, M. and {Berihuete}, A. and {Bernet}, M. and {Bertone}, S. and {Bianchi}, L. and {Binnenfeld}, A. and {Blanco-Cuaresma}, S. and {Blazere}, A. and {Boch}, T. and {Bombrun}, A. and {Bossini}, D. and {Bouquillon}, S. and {Bragaglia}, A. and {Bramante}, L. and {Breedt}, E. and {Bressan}, A. and {Brouillet}, N. and {Brugaletta}, E. and {Bucciarelli}, B. and {Burlacu}, A. and {Butkevich}, A.~G. and {Buzzi}, R. and {Caffau}, E. and {Cancelliere}, R. and {Cantat-Gaudin}, T. and {Carballo}, R. and {Carlucci}, T. and {Carnerero}, M.~I. and {Carrasco}, J.~M. and {Casamiquela}, L. and {Castellani}, M. and {Castro-Ginard}, A. and {Chaoul}, L. and {Charlot}, P. and {Chemin}, L. and {Chiaramida}, V. and {Chiavassa}, A. and {Chornay}, N. and {Comoretto}, G. and {Contursi}, G. and {Cooper}, W.~J. and {Cornez}, T. and {Cowell}, S. and {Crifo}, F. and {Cropper}, M. and {Crosta}, M. and {Crowley}, C. and {Dafonte}, C. and {Dapergolas}, A. and {David}, M. and {David}, P. and {de Laverny}, P. and {De Luise}, F. and {De March}, R. and {De Ridder}, J. and {de Souza}, R. and {de Torres}, A. and {del Peloso}, E.~F. and {del Pozo}, E. and {Delbo}, M. and {Delgado}, A. and {Delisle}, J. -B. and {Demouchy}, C. and {Dharmawardena}, T.~E. and {Di Matteo}, P. and {Diakite}, S. and {Diener}, C. and {Distefano}, E. and {Dolding}, C. and {Edvardsson}, B. and {Enke}, H. and {Fabre}, C. and {Fabrizio}, M. and {Faigler}, S. and {Fedorets}, G. and {Fernique}, P. and {Fienga}, A. and {Figueras}, F. and {Fournier}, Y. and {Fouron}, C. and {Fragkoudi}, F. and {Gai}, M. and {Garcia-Gutierrez}, A. and {Garcia-Reinaldos}, M. and {Garc{\'\i}a-Torres}, M. and {Garofalo}, A. and {Gavel}, A. and {Gavras}, P. and {Gerlach}, E. and {Geyer}, R. and {Giacobbe}, P. and {Gilmore}, G. and {Girona}, S. and {Giuffrida}, G. and {Gomel}, R. and {Gomez}, A. and {Gonz{\'a}lez-N{\'u}{\~n}ez}, J. and {Gonz{\'a}lez-Santamar{\'\i}a}, I. and {Gonz{\'a}lez-Vidal}, J.~J. and {Granvik}, M. and {Guillout}, P. and {Guiraud}, J. and {Guti{\'e}rrez-S{\'a}nchez}, R. and {Guy}, L.~P. and {Hatzidimitriou}, D. and {Hauser}, M. and {Haywood}, M. and {Helmer}, A. and {Helmi}, A. and {Sarmiento}, M.~H. and {Hidalgo}, S.~L. and {Hilger}, T. and {H{\l}adczuk}, N. and {Hobbs}, D. and {Holland}, G. and {Huckle}, H.~E. and {Jardine}, K. and {Jasniewicz}, G. and {Jean-Antoine Piccolo}, A. and {Jim{\'e}nez-Arranz}, {\'O}. and {Jorissen}, A. and {Juaristi Campillo}, J. and {Julbe}, F. and {Karbevska}, L. and {Kervella}, P. and {Khanna}, S. and {Kontizas}, M. and {Kordopatis}, G. and {Korn}, A.~J. and {K{\'o}sp{\'a}l}, {\'A}. and {Kostrzewa-Rutkowska}, Z. and {Kruszy{\'n}ska}, K. and {Kun}, M. and {Laizeau}, P. and {Lambert}, S. and {Lanza}, A.~F. and {Lasne}, Y. and {Le Campion}, J. -F. and {Lebreton}, Y. and {Lebzelter}, T. and {Leccia}, S. and {Leclerc}, N. and {Lecoeur-Taibi}, I. and {Liao}, S. and {Licata}, E.~L. and {Lindstr{\o}m}, H.~E.~P. and {Lister}, T.~A. and {Livanou}, E. and {Lobel}, A. and {Lorca}, A. and {Loup}, C. and {Madrero Pardo}, P. and {Magdaleno Romeo}, A. and {Managau}, S. and {Mann}, R.~G. and {Manteiga}, M. and {Marchant}, J.~M. and {Marconi}, M. and {Marcos}, J. and {Marcos Santos}, M.~M.~S. and {Mar{\'\i}n Pina}, D. and {Marinoni}, S. and {Marocco}, F. and {Marshall}, D.~J. and {Martin Polo}, L. and {Mart{\'\i}n-Fleitas}, J.~M. and {Marton}, G. and {Mary}, N. and {Masip}, A. and {Massari}, D. and {Mastrobuono-Battisti}, A. and {Mazeh}, T. and {McMillan}, P.~J. and {Messina}, S. and {Michalik}, D. and {Millar}, N.~R. and {Mints}, A. and {Molina}, D. and {Molinaro}, R. and {Moln{\'a}r}, L. and {Monari}, G. and {Mongui{\'o}}, M. and {Montegriffo}, P. and {Montero}, A. and {Mor}, R. and {Mora}, A. and {Morbidelli}, R. and {Morel}, T. and {Morris}, D. and {Muraveva}, T. and {Murphy}, C.~P. and {Musella}, I. and {Nagy}, Z. and {Noval}, L. and {Oca{\~n}a}, F. and {Ogden}, A. and {Ordenovic}, C. and {Osinde}, J.~O. and {Pagani}, C. and {Pagano}, I. and {Palaversa}, L. and {Palicio}, P.~A. and {Pallas-Quintela}, L. and {Panahi}, A. and {Payne-Wardenaar}, S. and {Pe{\~n}alosa Esteller}, X. and {Penttil{\"a}}, A. and {Pichon}, B. and {Piersimoni}, A.~M. and {Pineau}, F. -X. and {Plachy}, E. and {Plum}, G. and {Poggio}, E. and {Pr{\v{s}}a}, A. and {Pulone}, L. and {Racero}, E. and {Ragaini}, S. and {Rainer}, M. and {Raiteri}, C.~M. and {Rambaux}, N. and {Ramos}, P. and {Ramos-Lerate}, M. and {Re Fiorentin}, P. and {Regibo}, S. and {Richards}, P.~J. and {Rios Diaz}, C. and {Ripepi}, V. and {Riva}, A. and {Rix}, H. -W. and {Rixon}, G. and {Robichon}, N. and {Robin}, A.~C. and {Robin}, C. and {Roelens}, M. and {Rogues}, H.~R.~O. and {Rohrbasser}, L. and {Romero-G{\'o}mez}, M. and {Rowell}, N. and {Royer}, F. and {Ruz Mieres}, D. and {Rybicki}, K.~A. and {Sadowski}, G. and {S{\'a}ez N{\'u}{\~n}ez}, A. and {Sagrist{\`a} Sell{\'e}s}, A. and {Sahlmann}, J. and {Salguero}, E. and {Samaras}, N. and {Sanchez Gimenez}, V. and {Sanna}, N. and {Santove{\~n}a}, R. and {Sarasso}, M. and {Schultheis}, M. and {Sciacca}, E. and {Segol}, M. and {Segovia}, J.~C. and {S{\'e}gransan}, D. and {Semeux}, D. and {Shahaf}, S. and {Siddiqui}, H.~I. and {Siebert}, A. and {Siltala}, L. and {Silvelo}, A. and {Slezak}, E. and {Slezak}, I. and {Smart}, R.~L. and {Snaith}, O.~N. and {Solano}, E. and {Solitro}, F. and {Souami}, D. and {Souchay}, J. and {Spagna}, A. and {Spina}, L. and {Spoto}, F. and {Steele}, I.~A. and {Steidelm{\"u}ller}, H. and {Stephenson}, C.~A. and {S{\"u}veges}, M. and {Surdej}, J. and {Szabados}, L. and {Szegedi-Elek}, E. and {Taris}, F. and {Taylor}, M.~B. and {Teixeira}, R. and {Tolomei}, L. and {Tonello}, N. and {Torra}, F. and {Torra}, J. and {Torralba Elipe}, G. and {Trabucchi}, M. and {Tsounis}, A.~T. and {Turon}, C. and {Ulla}, A. and {Unger}, N. and {Vaillant}, M.~V. and {van Dillen}, E. and {van Reeven}, W. and {Vanel}, O. and {Vecchiato}, A. and {Viala}, Y. and {Vicente}, D. and {Voutsinas}, S. and {Weiler}, M. and {Wevers}, T. and {Wyrzykowski}, {\L}. and {Yoldas}, A. and {Yvard}, P. and {Zhao}, H. and {Zorec}, J. and {Zucker}, S. and {Zwitter}, T.},
        title = "{Gaia Data Release 3. Summary of the content and survey properties}",
      journal = {\aap},
         year = 2023,
        month = jun,
       volume = {674},
          eid = {A1},
        pages = {A1},
          doi = {10.1051/0004-6361/202243940},
archivePrefix = {arXiv},
       eprint = {2208.00211},
 primaryClass = {astro-ph.GA},
       adsurl = {https://ui.adsabs.harvard.edu/abs/2023A&A...674A...1G}
}

@ARTICLE{_naidy20,
       author = {{Naidu}, Rohan P. and {Conroy}, Charlie and {Bonaca}, Ana and {Johnson}, Benjamin D. and {Ting}, Yuan-Sen and {Caldwell}, Nelson and {Zaritsky}, Dennis and {Cargile}, Phillip A.},
        title = "{Evidence from the H3 Survey That the Stellar Halo Is Entirely Comprised of Substructure}",
      journal = {\apj},
         year = 2020,
        month = sep,
       volume = {901},
       number = {1},
          eid = {48},
        pages = {48},
          doi = {10.3847/1538-4357/abaef4},
archivePrefix = {arXiv},
       eprint = {2006.08625},
 primaryClass = {astro-ph.GA},
       adsurl = {https://ui.adsabs.harvard.edu/abs/2020ApJ...901...48N}
}

@ARTICLE{_buder21,
       author = {{Buder}, Sven and {Sharma}, Sanjib and {Kos}, Janez and {Amarsi}, Anish M. and {Nordlander}, Thomas and {Lind}, Karin and {Martell}, Sarah L. and {Asplund}, Martin and {Bland-Hawthorn}, Joss and {Casey}, Andrew R. and {de Silva}, Gayandhi M. and {D'Orazi}, Valentina and {Freeman}, Ken C. and {Hayden}, Michael R. and {Lewis}, Geraint F. and {Lin}, Jane and {Schlesinger}, Katharine J. and {Simpson}, Jeffrey D. and {Stello}, Dennis and {Zucker}, Daniel B. and {Zwitter}, Toma{\v{z}} and {Beeson}, Kevin L. and {Buck}, Tobias and {Casagrande}, Luca and {Clark}, Jake T. and {{\v{C}}otar}, Klemen and {da Costa}, Gary S. and {de Grijs}, Richard and {Feuillet}, Diane and {Horner}, Jonathan and {Kafle}, Prajwal R. and {Khanna}, Shourya and {Kobayashi}, Chiaki and {Liu}, Fan and {Montet}, Benjamin T. and {Nandakumar}, Govind and {Nataf}, David M. and {Ness}, Melissa K. and {Spina}, Lorenzo and {Tepper-Garc{\'\i}a}, Thor and {Ting}, Yuan-Sen and {Traven}, Gregor and {Vogrin{\v{c}}i{\v{c}}}, Rok and {Wittenmyer}, Robert A. and {Wyse}, Rosemary F.~G. and {{\v{Z}}erjal}, Maru{\v{s}}a and {GALAH Collaboration}},
        title = "{The GALAH+ survey: Third data release}",
      journal = {\mnras},
         year = 2021,
        month = sep,
       volume = {506},
       number = {1},
        pages = {150-201},
          doi = {10.1093/mnras/stab1242},
archivePrefix = {arXiv},
       eprint = {2011.02505},
 primaryClass = {astro-ph.GA},
       adsurl = {https://ui.adsabs.harvard.edu/abs/2021MNRAS.506..150B}
}

@ARTICLE{_belokurov20,
       author = {{Belokurov}, Vasily and {Sanders}, Jason L. and {Fattahi}, Azadeh and {Smith}, Martin C. and {Deason}, Alis J. and {Evans}, N. Wyn and {Grand}, Robert J.~J.},
        title = "{The biggest splash}",
      journal = {\mnras},
         year = 2020,
        month = may,
       volume = {494},
       number = {3},
        pages = {3880-3898},
          doi = {10.1093/mnras/staa876},
archivePrefix = {arXiv},
       eprint = {1909.04679},
 primaryClass = {astro-ph.GA},
       adsurl = {https://ui.adsabs.harvard.edu/abs/2020MNRAS.494.3880B}
}

@ARTICLE{_pywt,
       author = {{Lee}, Gregory and {Gommers}, Ralf and {Waselewski}, Filip and {Wohlfahrt}, Kai and {O'Leary}, Aaron},
        title = "{PyWavelets: A Python package for wavelet analysis}",
      journal = {The Journal of Open Source Software},
         year = 2019,
        month = apr,
       volume = {4},
       number = {36},
          eid = {1237},
        pages = {1237},
          doi = {10.21105/joss.01237},
       adsurl = {https://ui.adsabs.harvard.edu/abs/2019JOSS....4.1237L}
}

@ARTICLE{_galpy,
       author = {{Bovy}, Jo},
        title = "{galpy: A python Library for Galactic Dynamics}",
      journal = {\apjs},
         year = 2015,
        month = feb,
       volume = {216},
       number = {2},
          eid = {29},
        pages = {29},
          doi = {10.1088/0067-0049/216/2/29},
archivePrefix = {arXiv},
       eprint = {1412.3451},
 primaryClass = {astro-ph.GA},
       adsurl = {https://ui.adsabs.harvard.edu/abs/2015ApJS..216...29B}
}

@ARTICLE{_contursi23,
       author = {{Contursi}, G. and {de Laverny}, P. and {Recio-Blanco}, A. and {Spitoni}, E. and {Palicio}, P.~A. and {Poggio}, E. and {Grisoni}, V. and {Cescutti}, G. and {Matteucci}, F. and {Spina}, L. and {{\'A}lvarez}, M.~A. and {Kordopatis}, G. and {Ordenovic}, C. and {Oreshina-Slezak}, I. and {Zhao}, H.},
        title = "{The cerium content of the Milky Way as revealed by Gaia DR3 GSP-Spec abundances}",
      journal = {\aap},
         year = 2023,
        month = feb,
       volume = {670},
          eid = {A106},
        pages = {A106},
          doi = {10.1051/0004-6361/202244469},
archivePrefix = {arXiv},
       eprint = {2207.05368},
 primaryClass = {astro-ph.GA},
       adsurl = {https://ui.adsabs.harvard.edu/abs/2023A&A...670A.106C}
}

@ARTICLE{_myeong19,
       author = {{Myeong}, G.~C. and {Vasiliev}, E. and {Iorio}, G. and {Evans}, N.~W. and {Belokurov}, V.},
        title = "{Evidence for two early accretion events that built the Milky Way stellar halo}",
      journal = {\mnras},
         year = 2019,
        month = sep,
       volume = {488},
       number = {1},
        pages = {1235-1247},
          doi = {10.1093/mnras/stz1770},
archivePrefix = {arXiv},
       eprint = {1904.03185},
 primaryClass = {astro-ph.GA},
       adsurl = {https://ui.adsabs.harvard.edu/abs/2019MNRAS.488.1235M}
}

@ARTICLE{_kobayashi20,
       author = {{Kobayashi}, Chiaki and {Karakas}, Amanda I. and {Lugaro}, Maria},
        title = "{The Origin of Elements from Carbon to Uranium}",
      journal = {\apj},
         year = 2020,
        month = sep,
       volume = {900},
       number = {2},
          eid = {179},
        pages = {179},
          doi = {10.3847/1538-4357/abae65},
archivePrefix = {arXiv},
       eprint = {2008.04660},
 primaryClass = {astro-ph.GA},
       adsurl = {https://ui.adsabs.harvard.edu/abs/2020ApJ...900..179K}
}

@ARTICLE{_s5_20,
       author = {{Ji}, Alexander P. and {Li}, Ting S. and {Hansen}, Terese T. and {Casey}, Andrew R. and {Koposov}, Sergey E. and {Pace}, Andrew B. and {Mackey}, Dougal and {Lewis}, Geraint F. and {Simpson}, Jeffrey D. and {Bland-Hawthorn}, Joss and {Cullinane}, Lara R. and {Da Costa}, Gary. S. and {Hattori}, Kohei and {Martell}, Sarah L. and {Kuehn}, Kyler and {Erkal}, Denis and {Shipp}, Nora and {Wan}, Zhen and {Zucker}, Daniel B.},
        title = "{The Southern Stellar Stream Spectroscopic Survey (S$^{5}$): Chemical Abundances of Seven Stellar Streams}",
      journal = {\aj},
         year = 2020,
        month = oct,
       volume = {160},
       number = {4},
          eid = {181},
        pages = {181},
          doi = {10.3847/1538-3881/abacb6},
archivePrefix = {arXiv},
       eprint = {2008.07568},
 primaryClass = {astro-ph.SR},
       adsurl = {https://ui.adsabs.harvard.edu/abs/2020AJ....160..181J}
}

@ARTICLE{_nissen24,
       author = {{Nissen}, P.~E. and {Amarsi}, A.~M. and {Sk{\'u}lad{\'o}ttir}, {\'A}. and {Schuster}, W.~J.},
        title = "{Abundances of iron-peak elements in accreted and in situ born Galactic halo stars}",
      journal = {\aap},
         year = 2024,
        month = feb,
       volume = {682},
          eid = {A116},
        pages = {A116},
          doi = {10.1051/0004-6361/202348392},
archivePrefix = {arXiv},
       eprint = {2312.07768},
 primaryClass = {astro-ph.GA},
       adsurl = {https://ui.adsabs.harvard.edu/abs/2024A&A...682A.116N}
}

@ARTICLE{_abdurro22,
       author = {{Abdurro'uf} and {Accetta}, Katherine and {Aerts}, Conny and {Silva Aguirre}, V{\'\i}ctor and {Ahumada}, Romina and {Ajgaonkar}, Nikhil and {Filiz Ak}, N. and {Alam}, Shadab and {Allende Prieto}, Carlos and {Almeida}, Andr{\'e}s and {Anders}, Friedrich and {Anderson}, Scott F. and {Andrews}, Brett H. and {Anguiano}, Borja and {Aquino-Ort{\'\i}z}, Erik and {Arag{\'o}n-Salamanca}, Alfonso and {Argudo-Fern{\'a}ndez}, Maria and {Ata}, Metin and {Aubert}, Marie and {Avila-Reese}, Vladimir and {Badenes}, Carles and {Barb{\'a}}, Rodolfo H. and {Barger}, Kat and {Barrera-Ballesteros}, Jorge K. and {Beaton}, Rachael L. and {Beers}, Timothy C. and {Belfiore}, Francesco and {Bender}, Chad F. and {Bernardi}, Mariangela and {Bershady}, Matthew A. and {Beutler}, Florian and {Bidin}, Christian Moni and {Bird}, Jonathan C. and {Bizyaev}, Dmitry and {Blanc}, Guillermo A. and {Blanton}, Michael R. and {Boardman}, Nicholas Fraser and {Bolton}, Adam S. and {Boquien}, M{\'e}d{\'e}ric and {Borissova}, Jura and {Bovy}, Jo and {Brandt}, W.~N. and {Brown}, Jordan and {Brownstein}, Joel R. and {Brusa}, Marcella and {Buchner}, Johannes and {Bundy}, Kevin and {Burchett}, Joseph N. and {Bureau}, Martin and {Burgasser}, Adam and {Cabang}, Tuesday K. and {Campbell}, Stephanie and {Cappellari}, Michele and {Carlberg}, Joleen K. and {Wanderley}, F{\'a}bio Carneiro and {Carrera}, Ricardo and {Cash}, Jennifer and {Chen}, Yan-Ping and {Chen}, Wei-Huai and {Cherinka}, Brian and {Chiappini}, Cristina and {Choi}, Peter Doohyun and {Chojnowski}, S. Drew and {Chung}, Haeun and {Clerc}, Nicolas and {Cohen}, Roger E. and {Comerford}, Julia M. and {Comparat}, Johan and {da Costa}, Luiz and {Covey}, Kevin and {Crane}, Jeffrey D. and {Cruz-Gonzalez}, Irene and {Culhane}, Connor and {Cunha}, Katia and {Dai}, Y. Sophia and {Damke}, Guillermo and {Darling}, Jeremy and {Davidson}, James W., Jr. and {Davies}, Roger and {Dawson}, Kyle and {De Lee}, Nathan and {Diamond-Stanic}, Aleksandar M. and {Cano-D{\'\i}az}, Mariana and {S{\'a}nchez}, Helena Dom{\'\i}nguez and {Donor}, John and {Duckworth}, Chris and {Dwelly}, Tom and {Eisenstein}, Daniel J. and {Elsworth}, Yvonne P. and {Emsellem}, Eric and {Eracleous}, Mike and {Escoffier}, Stephanie and {Fan}, Xiaohui and {Farr}, Emily and {Feng}, Shuai and {Fern{\'a}ndez-Trincado}, Jos{\'e} G. and {Feuillet}, Diane and {Filipp}, Andreas and {Fillingham}, Sean P. and {Frinchaboy}, Peter M. and {Fromenteau}, Sebastien and {Galbany}, Llu{\'\i}s and {Garc{\'\i}a}, Rafael A. and {Garc{\'\i}a-Hern{\'a}ndez}, D.~A. and {Ge}, Junqiang and {Geisler}, Doug and {Gelfand}, Joseph and {G{\'e}ron}, Tobias and {Gibson}, Benjamin J. and {Goddy}, Julian and {Godoy-Rivera}, Diego and {Grabowski}, Kathleen and {Green}, Paul J. and {Greener}, Michael and {Grier}, Catherine J. and {Griffith}, Emily and {Guo}, Hong and {Guy}, Julien and {Hadjara}, Massinissa and {Harding}, Paul and {Hasselquist}, Sten and {Hayes}, Christian R. and {Hearty}, Fred and {Hern{\'a}ndez}, Jes{\'u}s and {Hill}, Lewis and {Hogg}, David W. and {Holtzman}, Jon A. and {Horta}, Danny and {Hsieh}, Bau-Ching and {Hsu}, Chin-Hao and {Hsu}, Yun-Hsin and {Huber}, Daniel and {Huertas-Company}, Marc and {Hutchinson}, Brian and {Hwang}, Ho Seong and {Ibarra-Medel}, H{\'e}ctor J. and {Chitham}, Jacob Ider and {Ilha}, Gabriele S. and {Imig}, Julie and {Jaekle}, Will and {Jayasinghe}, Tharindu and {Ji}, Xihan and {Johnson}, Jennifer A. and {Jones}, Amy and {J{\"o}nsson}, Henrik and {Katkov}, Ivan and {Khalatyan}, Arman, Dr. and {Kinemuchi}, Karen and {Kisku}, Shobhit and {Knapen}, Johan H. and {Kneib}, Jean-Paul and {Kollmeier}, Juna A. and {Kong}, Miranda and {Kounkel}, Marina and {Kreckel}, Kathryn and {Krishnarao}, Dhanesh and {Lacerna}, Ivan and {Lane}, Richard R. and {Langgin}, Rachel and {Lavender}, Ramon and {Law}, David R. and {Lazarz}, Daniel and {Leung}, Henry W. and {Leung}, Ho-Hin and {Lewis}, Hannah M. and {Li}, Cheng and {Li}, Ran and {Lian}, Jianhui and {Liang}, Fu-Heng and {Lin}, Lihwai and {Lin}, Yen-Ting and {Lin}, Sicheng and {Lintott}, Chris and {Long}, Dan and {Longa-Pe{\~n}a}, Pen{\'e}lope and {L{\'o}pez-Cob{\'a}}, Carlos and {Lu}, Shengdong and {Lundgren}, Britt F. and {Luo}, Yuanze and {Mackereth}, J. Ted and {de la Macorra}, Axel and {Mahadevan}, Suvrath and {Majewski}, Steven R. and {Manchado}, Arturo and {Mandeville}, Travis and {Maraston}, Claudia and {Margalef-Bentabol}, Berta and {Masseron}, Thomas and {Masters}, Karen L. and {Mathur}, Savita and {McDermid}, Richard M. and {Mckay}, Myles and {Merloni}, Andrea and {Merrifield}, Michael and {Meszaros}, Szabolcs and {Miglio}, Andrea and {Di Mille}, Francesco and {Minniti}, Dante and {Minsley}, Rebecca and {Monachesi}, Antonela and {Moon}, Jeongin and {Mosser}, Benoit and {Mulchaey}, John and {Muna}, Demitri and {Mu{\~n}oz}, Ricardo R. and {Myers}, Adam D. and {Myers}, Natalie and {Nadathur}, Seshadri and {Nair}, Preethi and {Nandra}, Kirpal and {Neumann}, Justus and {Newman}, Jeffrey A. and {Nidever}, David L. and {Nikakhtar}, Farnik and {Nitschelm}, Christian and {O'Connell}, Julia E. and {Garma-Oehmichen}, Luis and {Luan Souza de Oliveira}, Gabriel and {Olney}, Richard and {Oravetz}, Daniel and {Ortigoza-Urdaneta}, Mario and {Osorio}, Yeisson and {Otter}, Justin and {Pace}, Zachary J. and {Padilla}, Nelson and {Pan}, Kaike and {Pan}, Hsi-An and {Parikh}, Taniya and {Parker}, James and {Peirani}, Sebastien and {Pe{\~n}a Ram{\'\i}rez}, Karla and {Penny}, Samantha and {Percival}, Will J. and {Perez-Fournon}, Ismael and {Pinsonneault}, Marc and {Poidevin}, Fr{\'e}d{\'e}rick and {Poovelil}, Vijith Jacob and {Price-Whelan}, Adrian M. and {B{\'a}rbara de Andrade Queiroz}, Anna and {Raddick}, M. Jordan and {Ray}, Amy and {Rembold}, Sandro Barboza and {Riddle}, Nicole and {Riffel}, Rogemar A. and {Riffel}, Rog{\'e}rio and {Rix}, Hans-Walter and {Robin}, Annie C. and {Rodr{\'\i}guez-Puebla}, Aldo and {Roman-Lopes}, Alexandre and {Rom{\'a}n-Z{\'u}{\~n}iga}, Carlos and {Rose}, Benjamin and {Ross}, Ashley J. and {Rossi}, Graziano and {Rubin}, Kate H.~R. and {Salvato}, Mara and {S{\'a}nchez}, Seb{\'a}stian F. and {S{\'a}nchez-Gallego}, Jos{\'e} R. and {Sanderson}, Robyn and {Santana Rojas}, Felipe Antonio and {Sarceno}, Edgar and {Sarmiento}, Regina and {Sayres}, Conor and {Sazonova}, Elizaveta and {Schaefer}, Adam L. and {Schiavon}, Ricardo and {Schlegel}, David J. and {Schneider}, Donald P. and {Schultheis}, Mathias and {Schwope}, Axel and {Serenelli}, Aldo and {Serna}, Javier and {Shao}, Zhengyi and {Shapiro}, Griffin and {Sharma}, Anubhav and {Shen}, Yue and {Shetrone}, Matthew and {Shu}, Yiping and {Simon}, Joshua D. and {Skrutskie}, M.~F. and {Smethurst}, Rebecca and {Smith}, Verne and {Sobeck}, Jennifer and {Spoo}, Taylor and {Sprague}, Dani and {Stark}, David V. and {Stassun}, Keivan G. and {Steinmetz}, Matthias and {Stello}, Dennis and {Stone-Martinez}, Alexander and {Storchi-Bergmann}, Thaisa and {Stringfellow}, Guy S. and {Stutz}, Amelia and {Su}, Yung-Chau and {Taghizadeh-Popp}, Manuchehr and {Talbot}, Michael S. and {Tayar}, Jamie and {Telles}, Eduardo and {Teske}, Johanna and {Thakar}, Ani and {Theissen}, Christopher and {Tkachenko}, Andrew and {Thomas}, Daniel and {Tojeiro}, Rita and {Hernandez Toledo}, Hector and {Troup}, Nicholas W. and {Trump}, Jonathan R. and {Trussler}, James and {Turner}, Jacqueline and {Tuttle}, Sarah and {Unda-Sanzana}, Eduardo and {V{\'a}zquez-Mata}, Jos{\'e} Antonio and {Valentini}, Marica and {Valenzuela}, Octavio and {Vargas-Gonz{\'a}lez}, Jaime and {Vargas-Maga{\~n}a}, Mariana and {Alfaro}, Pablo Vera and {Villanova}, Sandro and {Vincenzo}, Fiorenzo and {Wake}, David and {Warfield}, Jack T. and {Washington}, Jessica Diane and {Weaver}, Benjamin Alan and {Weijmans}, Anne-Marie and {Weinberg}, David H. and {Weiss}, Achim and {Westfall}, Kyle B. and {Wild}, Vivienne and {Wilde}, Matthew C. and {Wilson}, John C. and {Wilson}, Robert F. and {Wilson}, Mikayla and {Wolf}, Julien and {Wood-Vasey}, W.~M. and {Yan}, Renbin and {Zamora}, Olga and {Zasowski}, Gail and {Zhang}, Kai and {Zhao}, Cheng and {Zheng}, Zheng and {Zheng}, Zheng and {Zhu}, Kai},
        title = "{The Seventeenth Data Release of the Sloan Digital Sky Surveys: Complete Release of MaNGA, MaStar, and APOGEE-2 Data}",
      journal = {\apjs},
         year = 2022,
        month = apr,
       volume = {259},
       number = {2},
          eid = {35},
        pages = {35},
          doi = {10.3847/1538-4365/ac4414},
archivePrefix = {arXiv},
       eprint = {2112.02026},
 primaryClass = {astro-ph.GA},
       adsurl = {https://ui.adsabs.harvard.edu/abs/2022ApJS..259...35A}
}

@ARTICLE{_sanders21,
       author = {{Sanders}, Jason L. and {Belokurov}, Vasily and {Man}, Kai T.~F.},
        title = "{Evidence for sub-Chandrasekhar Type Ia supernovae from the last major merger}",
      journal = {\mnras},
         year = 2021,
        month = sep,
       volume = {506},
       number = {3},
        pages = {4321-4343},
          doi = {10.1093/mnras/stab1951},
archivePrefix = {arXiv},
       eprint = {2106.11324},
 primaryClass = {astro-ph.GA},
       adsurl = {https://ui.adsabs.harvard.edu/abs/2021MNRAS.506.4321S}
}

@ARTICLE{_bellazzini23,
       author = {{Bellazzini}, M. and {Massari}, D. and {De Angeli}, F. and {Mucciarelli}, A. and {Bragaglia}, A. and {Riello}, M. and {Montegriffo}, P.},
        title = "{Photometric metallicity for 694 233 Galactic giant stars from Gaia DR3 synthetic Str{\"o}mgren photometry. Metallicity distribution functions of halo substructures}",
      journal = {\aap},
         year = 2023,
        month = jun,
       volume = {674},
          eid = {A194},
        pages = {A194},
          doi = {10.1051/0004-6361/202345921},
archivePrefix = {arXiv},
       eprint = {2304.10772},
 primaryClass = {astro-ph.GA},
       adsurl = {https://ui.adsabs.harvard.edu/abs/2023A&A...674A.194B}
}

@ARTICLE{_ortigoza23,
       author = {{Ortigoza-Urdaneta}, Mario and {Vieira}, Katherine and {Fern{\'a}ndez-Trincado}, Jos{\'e} G. and {Queiroz}, Anna B.~A. and {Barbuy}, Beatriz and {Beers}, Timothy C. and {Chiappini}, Cristina and {Anders}, Friedrich and {Minniti}, Dante and {Tang}, Baitian},
        title = "{Galactic ArchaeoLogIcaL ExcavatiOns (GALILEO). II. t-SNE portrait of local fossil relics and structures}",
      journal = {\aap},
         year = 2023,
        month = aug,
       volume = {676},
          eid = {A140},
        pages = {A140},
          doi = {10.1051/0004-6361/202346325},
archivePrefix = {arXiv},
       eprint = {2306.08677},
 primaryClass = {astro-ph.GA},
       adsurl = {https://ui.adsabs.harvard.edu/abs/2023A&A...676A.140O}
}

@ARTICLE{_conroy19,
       author = {{Conroy}, Charlie and {Bonaca}, Ana and {Cargile}, Phillip and {Johnson}, Benjamin D. and {Caldwell}, Nelson and {Naidu}, Rohan P. and {Zaritsky}, Dennis and {Fabricant}, Daniel and {Moran}, Sean and {Rhee}, Jaehyon and {Szentgyorgyi}, Andrew and {Berlind}, Perry and {Calkins}, Michael L. and {Kattner}, ShiAnne and {Ly}, Chun},
        title = "{Mapping the Stellar Halo with the H3 Spectroscopic Survey}",
      journal = {\apj},
         year = 2019,
        month = sep,
       volume = {883},
       number = {1},
          eid = {107},
        pages = {107},
          doi = {10.3847/1538-4357/ab38b8},
archivePrefix = {arXiv},
       eprint = {1907.07684},
 primaryClass = {astro-ph.GA},
       adsurl = {https://ui.adsabs.harvard.edu/abs/2019ApJ...883..107C}
}

@ARTICLE{_haywood18,
       author = {{Haywood}, M. and {Di Matteo}, P. and {Lehnert}, M.~D. and {Snaith}, O. and {Khoperskov}, S. and {G{\'o}mez}, A.},
        title = "{In Disguise or Out of Reach: First Clues about In Situ and Accreted Stars in the Stellar Halo of the Milky Way from Gaia DR2}",
      journal = {\apj},
         year = 2018,
        month = aug,
       volume = {863},
       number = {2},
          eid = {113},
        pages = {113},
          doi = {10.3847/1538-4357/aad235},
archivePrefix = {arXiv},
       eprint = {1805.02617},
 primaryClass = {astro-ph.GA},
       adsurl = {https://ui.adsabs.harvard.edu/abs/2018ApJ...863..113H}
}

@ARTICLE{_bailerjones23,
       author = {{Bailer-Jones}, C.~A.~L.},
        title = "{Estimating Distances from Parallaxes. VI. A Method for Inferring Distances and Transverse Velocities from Parallaxes and Proper Motions Demonstrated on Gaia Data Release 3}",
      journal = {\aj},
         year = 2023,
        month = dec,
       volume = {166},
       number = {6},
          eid = {269},
        pages = {269},
          doi = {10.3847/1538-3881/ad08bb},
archivePrefix = {arXiv},
       eprint = {2311.00374},
 primaryClass = {astro-ph.GA},
       adsurl = {https://ui.adsabs.harvard.edu/abs/2023AJ....166..269B}
}

@ARTICLE{_deason24,
       author = {{Deason}, Alis J. and {Belokurov}, Vasily},
        title = "{Galactic Archaeology with Gaia}",
      journal = {\nar},
         year = 2024,
        month = dec,
       volume = {99},
          eid = {101706},
        pages = {101706},
          doi = {10.1016/j.newar.2024.101706},
archivePrefix = {arXiv},
       eprint = {2402.12443},
 primaryClass = {astro-ph.GA},
       adsurl = {https://ui.adsabs.harvard.edu/abs/2024NewAR..9901706D}
}

@ARTICLE{_feuillet21,
       author = {{Feuillet}, Diane K. and {Sahlholdt}, Christian L. and {Feltzing}, Sofia and {Casagrande}, Luca},
        title = "{Selecting accreted populations: metallicity, elemental abundances, and ages of the Gaia-Sausage-Enceladus and Sequoia populations}",
      journal = {\mnras},
         year = 2021,
        month = nov,
       volume = {508},
       number = {1},
        pages = {1489-1508},
          doi = {10.1093/mnras/stab2614},
archivePrefix = {arXiv},
       eprint = {2105.12141},
 primaryClass = {astro-ph.GA},
       adsurl = {https://ui.adsabs.harvard.edu/abs/2021MNRAS.508.1489F}
}

@INPROCEEDINGS{_dalton18,
       author = {{Dalton}, Gavin and {Trager}, Scott and {Abrams}, Don Carlos and {Bonifacio}, Piercarlo and {Aguerri}, J. Alfonso L. and {Vallenari}, Antonella and {Middleton}, Kevin and {Benn}, Chris and {Dee}, Kevin and {Say{\`e}de}, Fr{\'e}d{\'e}ric and {Lewis}, Ian and {Pragt}, Johannes and {Pic{\'o}}, Sergio and {Walton}, Nicholas and {Rey}, Jeurg and {Allende Prieto}, Carlos and {Lhom{\'e}}, {\'E}milie and {Terrett}, David and {Brock}, Matthew and {Gilbert}, James and {Ridings}, Andy and {Verheijen}, Marc and {Tosh}, Ian and {Steele}, Iain and {Stuik}, Remko and {Kroes}, Gabby and {Tromp}, Neils and {Kragt}, Jan and {Lesman}, Dirk and {Mottram}, Chris and {Bates}, Stuart and {Gribbin}, Frank and {Burgal}, Jos{\'e} Alonso and {Herreros}, Jose Miguel and {Delgado}, Jose Miguel and {Martin}, Carlos and {Cano}, Diego and {Navarro}, Ram{\'o}n and {Irwin}, Mike and {Lewis}, James and {Gonzales Solares}, Eduardo and {O'Mahony}, Neil and {Bianco}, Andrea and {Zurita}, Christina and {ter Horst}, Rik and {Molinari}, Emilio and {Lodi}, Marcello and {Guerra}, Jos{\'e} and {Baruffolo}, Andrea and {Carrasco}, Esperanza and {Farkas}, Szigfrid and {Schallig}, Ellen and {Hill}, Vanessa and {Smith}, Dan and {Drew}, Janet and {Poggianti}, Bianca and {Pieri}, Mat and {Jin}, Shoko and {Dominquez Palmero}, Lilian and {Fari{\~n}a}, Cecilia and {Martin}, Adrian and {Worley}, Clare and {Murphy}, David and {Hidalgo}, Andrea and {Mignot}, Shan and {Bishop}, Georgia and {Guest}, Steve and {Elswijk}, Eddy and {de Haan}, Menno and {Hanenburg}, Hiddo and {Salasnich}, Bernardo and {Mayya}, Divakara and {Izazaga-P{\'e}rez}, Rafael and {Peralta de Arriba}, Luis},
        title = "{Construction progress of WEAVE: the next generation wide-field spectroscopy facility for the William Herschel Telescope}",
    booktitle = {Ground-based and Airborne Instrumentation for Astronomy VII},
         year = 2018,
       editor = {{Evans}, Christopher J. and {Simard}, Luc and {Takami}, Hideki},
       series = {Society of Photo-Optical Instrumentation Engineers (SPIE) Conference Series},
       volume = {10702},
        month = jul,
          eid = {107021B},
        pages = {107021B},
          doi = {10.1117/12.2312031},
       adsurl = {https://ui.adsabs.harvard.edu/abs/2018SPIE10702E..1BD}
}

@INPROCEEDINGS{_dalton20,
       author = {{Dalton}, Gavin and {Trager}, Scott and {Abrams}, Don Carlos and {Bonifacio}, Piercarlo and {Aguerri}, J. Alfonso L. and {Vallenari}, Antonella and {Bishop}, Georgia and {Middleton}, Kevin and {Benn}, Chris and {Dee}, Kevin and {Mignot}, Shan and {Lewis}, Ian and {Pragt}, Johannes and {Pico}, Sergio and {Walton}, Nicholas and {Rey}, Juerg and {Allende Prieto}, Carlos and {Lhom{\'e}}, Emilie and {Balcells}, Marc and {Terrett}, David and {Brock}, Matthew and {Ridings}, Andy and {Skvar{\v{c}}}, Jure and {Verheijen}, Marc and {Steele}, Iain and {Stuik}, Remko and {Kroes}, Gabby and {Tromp}, Neils and {Kragt}, Jan and {Lesman}, Dirk and {Mottram}, Chris and {Bates}, Stuart and {Gribbin}, Frank and {Burgal}, Jose Alonso and {Herreros}, Jos{\'e} Miguel and {Delgado}, Jos{\'e} Miguel and {Martin}, Carlos and {Cano}, Diego and {Navarro}, Ramon and {Irwin}, Mike and {Peralta de Arriba}, Luis and {O'Mahoney}, Neil and {Bianco}, Andrea and {Moleinezhad}, Alireza and {ter Horst}, Rik and {Molinari}, Emilio and {Lodi}, Marcello and {Guerra}, Jos{\'e} and {Baruffalo}, Andrea and {Carrasco}, Esperanza and {Farcas}, Szigfrid and {Schallig}, Ellen and {Hughes}, Sarah and {Hill}, Vanessa and {Smith}, Dan and {Drew}, Janet and {Poggianti}, Bianca and {Iovino}, Angela and {Pieri}, Mat and {Jin}, Shoko and {Dominguez Palmero}, Lillian and {Fari{\~n}a}, Cecilia and {Mart{\'\i}n}, Adrian and {Worley}, Clare and {Murphy}, David and {Guest}, Steve and {Morris}, Huw and {Elswijk}, Eddy and {de Haan}, Menno and {Hanenburg}, Hiddo and {Salasnich}, Bernardo and {Mayya}, Divakara and {Izazaga-P{\'e}rez}, Rafael and {Gafton}, Emanuel and {Caffau}, Elisabetta and {Horville}, David and {Paz Chinch{\'o}n}, Francisco and {Falcon-Barosso}, Jesus and {G{\"a}nsicke}, Boris and {San Juan}, Jose and {Hernandez}, Nauzet},
        title = "{Integration and early testing of WEAVE: the next-generation spectroscopy facility for the William Herschel Telescope}",
    booktitle = {Ground-based and Airborne Instrumentation for Astronomy VIII},
         year = 2020,
       editor = {{Evans}, Christopher J. and {Bryant}, Julia J. and {Motohara}, Kentaro},
       series = {Society of Photo-Optical Instrumentation Engineers (SPIE) Conference Series},
       volume = {11447},
        month = dec,
          eid = {1144714},
        pages = {1144714},
          doi = {10.1117/12.2561067},
       adsurl = {https://ui.adsabs.harvard.edu/abs/2020SPIE11447E..14D}
}

@ARTICLE{_feltzing23,
       author = {{Feltzing}, Sofia and {Feuillet}, Diane},
        title = "{The Metal-weak Milky Way Stellar Disk Hidden in the Gaia-Sausage-Enceladus Debris: The APOGEE DR17 View}",
      journal = {\apj},
         year = 2023,
        month = aug,
       volume = {953},
       number = {2},
          eid = {143},
        pages = {143},
          doi = {10.3847/1538-4357/ace185},
archivePrefix = {arXiv},
       eprint = {2303.00016},
 primaryClass = {astro-ph.GA},
       adsurl = {https://ui.adsabs.harvard.edu/abs/2023ApJ...953..143F}
}

@ARTICLE{_horta21,
       author = {{Horta}, Danny and {Schiavon}, Ricardo P. and {Mackereth}, J. Ted and {Pfeffer}, Joel and {Mason}, Andrew C. and {Kisku}, Shobhit and {Fragkoudi}, Francesca and {Allende Prieto}, Carlos and {Cunha}, Katia and {Hasselquist}, Sten and {Holtzman}, Jon and {Majewski}, Steven R. and {Nataf}, David and {O'Connell}, Robert W. and {Schultheis}, Mathias and {Smith}, Verne V.},
        title = "{Evidence from APOGEE for the presence of a major building block of the halo buried in the inner Galaxy}",
      journal = {\mnras},
         year = 2021,
        month = jan,
       volume = {500},
       number = {1},
        pages = {1385-1403},
          doi = {10.1093/mnras/staa2987},
archivePrefix = {arXiv},
       eprint = {2007.10374},
 primaryClass = {astro-ph.GA},
       adsurl = {https://ui.adsabs.harvard.edu/abs/2021MNRAS.500.1385H}
}

@ARTICLE{_belokurov18,
       author = {{Belokurov}, V. and {Erkal}, D. and {Evans}, N.~W. and {Koposov}, S.~E. and
         {Deason}, A.~J.},
        title = "{Co-formation of the disc and the stellar halo}",
      journal = {\mnras},
         year = "2018",
        month = "Jul",
       volume = {478},
       number = {1},
        pages = {611-619},
          doi = {10.1093/mnras/sty982},
archivePrefix = {arXiv},
       eprint = {1802.03414},
 primaryClass = {astro-ph.GA},
       adsurl = {https://ui.adsabs.harvard.edu/abs/2018MNRAS.478..611B}
}

@BOOK{_starck06,
       author = {{Starck}, Jean-Luc and {Murtagh}, Fionn},
        title = "{Astronomical Image and Data Analysis}",
         year = 2006,
          doi = {10.1007/978-3-540-33025-7},
       adsurl = {https://ui.adsabs.harvard.edu/abs/2006aida.book.....S}
}

@ARTICLE{_feuillet20,
       author = {{Feuillet}, Diane K. and {Feltzing}, Sofia and {Sahlholdt}, Christian L. and {Casagrande}, Luca},
        title = "{The SkyMapper-Gaia RVS view of the Gaia-Enceladus-Sausage - an investigation of the metallicity and mass of the Milky Way's last major merger}",
      journal = {\mnras},
         year = 2020,
        month = sep,
       volume = {497},
       number = {1},
        pages = {109-124},
          doi = {10.1093/mnras/staa1888},
archivePrefix = {arXiv},
       eprint = {2003.11039},
 primaryClass = {astro-ph.GA},
       adsurl = {https://ui.adsabs.harvard.edu/abs/2020MNRAS.497..109F}
}

@ARTICLE{_desilva15,
       author = {{De Silva}, G.~M. and {Freeman}, K.~C. and {Bland-Hawthorn}, J. and
         {Martell}, S. and {de Boer}, E. Wylie and {Asplund}, M. and
         {Keller}, S. and {Sharma}, S. and {Zucker}, D.~B. and {Zwitter}, T. and
         {Anguiano}, B. and {Bacigalupo}, C. and {Bayliss}, D. and
         {Beavis}, M.~A. and {Bergemann}, M. and {Campbell}, S. and
         {Cannon}, R. and {Carollo}, D. and {Casagrande}, L. and {Casey}, A.~R. and
         {Da Costa}, G. and {D'Orazi}, V. and {Dotter}, A. and {Duong}, L. and
         {Heger}, A. and {Ireland}, M.~J. and {Kafle}, P.~R. and {Kos}, J. and
         {Lattanzio}, J. and {Lewis}, G.~F. and {Lin}, J. and {Lind}, K. and
         {Munari}, U. and {Nataf}, D.~M. and {O'Toole}, S. and {Parker}, Q. and
         {Reid}, W. and {Schlesinger}, K.~J. and {Sheinis}, A. and
         {Simpson}, J.~D. and {Stello}, D. and {Ting}, Y. -S. and {Traven}, G. and
         {Watson}, F. and {Wittenmyer}, R. and {Yong}, D. and {{\v{Z}}erjal}, M.},
        title = "{The GALAH survey: scientific motivation}",
      journal = {\mnras},
         year = 2015,
        month = may,
       volume = {449},
       number = {3},
        pages = {2604-2617},
          doi = {10.1093/mnras/stv327},
archivePrefix = {arXiv},
       eprint = {1502.04767},
 primaryClass = {astro-ph.GA},
       adsurl = {https://ui.adsabs.harvard.edu/abs/2015MNRAS.449.2604D}
}

@ARTICLE{_helmi20,
       author = {{Helmi}, Amina},
        title = "{Streams, substructures and the early history of the Milky Way}",
      journal = {arXiv e-prints},
         year = 2020,
        month = feb,
          eid = {arXiv:2002.04340},
        pages = {arXiv:2002.04340},
archivePrefix = {arXiv},
       eprint = {2002.04340},
 primaryClass = {astro-ph.GA},
       adsurl = {https://ui.adsabs.harvard.edu/abs/2020arXiv200204340H}
}

@ARTICLE{_kushniruk19,
       author = {{Kushniruk}, Iryna and {Bensby}, Thomas},
        title = "{Disentangling the Arcturus stream}",
      journal = {\aap},
         year = 2019,
        month = nov,
       volume = {631},
          eid = {A47},
        pages = {A47},
          doi = {10.1051/0004-6361/201935234},
archivePrefix = {arXiv},
       eprint = {1909.04949},
 primaryClass = {astro-ph.GA},
       adsurl = {https://ui.adsabs.harvard.edu/abs/2019A&A...631A..47K}
}

@ARTICLE{_dejong2019,
       author = {{de Jong}, R.~S. and {Agertz}, O. and {Berbel}, A.~A. and {Aird}, J. and
         {Alexander}, D.~A. and {Amarsi}, A. and {Anders}, F. and {Andrae}, R. and
         {Ansarinejad}, B. and {Ansorge}, W. and {Antilogus}, P. and {Anwand
        -Heerwart}, H. and {Arentsen}, A. and {Arnadottir}, A. and
         {Asplund}, M. and {Auger}, M. and {Azais}, N. and {Baade}, D. and
         {Baker}, G. and {Baker}, S. and {Balbinot}, E. and {Baldry}, I.~K. and
         {Banerji}, M. and {Barden}, S. and {Barklem}, P. and
         {Barth{\'e}l{\'e}my-Mazot}, E. and {Battistini}, C. and {Bauer}, S. and
         {Bell}, C.~P.~M. and {Bellido-Tirado}, O. and {Bellstedt}, S. and
         {Belokurov}, V. and {Bensby}, T. and {Bergemann}, M. and
         {Bestenlehner}, J.~M. and {Bielby}, R. and {Bilicki}, M. and
         {Blake}, C. and {Bland-Hawthorn}, J. and {Boeche}, C. and {Boland}, W. and
         {Boller}, T. and {Bongard}, S. and {Bongiorno}, A. and {Bonifacio}, P. and
         {Boudon}, D. and {Brooks}, D. and {Brown}, M.~J.~I. and {Brown}, R. and
         {Br{\"u}ggen}, M. and {Brynnel}, J. and {Brzeski}, J. and
         {Buchert}, T. and {Buschkamp}, P. and {Caffau}, E. and {Caillier}, P. and
         {Carrick}, J. and {Casagrande}, L. and {Case}, S. and {Casey}, A. and
         {Cesarini}, I. and {Cescutti}, G. and {Chapuis}, D. and
         {Chiappini}, C. and {Childress}, M. and {Christlieb}, N. and
         {Church}, R. and {Cioni}, M. -R.~L. and {Cluver}, M. and {Colless}, M. and
         {Collett}, T. and {Comparat}, J. and {Cooper}, A. and {Couch}, W. and
         {Courbin}, F. and {Croom}, S. and {Croton}, D. and {Daguis{\'e}}, E. and
         {Dalton}, G. and {Davies}, L.~J.~M. and {Davis}, T. and
         {de Laverny}, P. and {Deason}, A. and {Dionies}, F. and {Disseau}, K. and
         {Doel}, P. and {D{\"o}scher}, D. and {Driver}, S.~P. and {Dwelly}, T. and
         {Eckert}, D. and {Edge}, A. and {Edvardsson}, B. and
         {Youssoufi}, D.~E. and {Elhaddad}, A. and {Enke}, H. and
         {Erfanianfar}, G. and {Farrell}, T. and {Fechner}, T. and {Feiz}, C. and
         {Feltzing}, S. and {Ferreras}, I. and {Feuerstein}, D. and
         {Feuillet}, D. and {Finoguenov}, A. and {Ford}, D. and
         {Fotopoulou}, S. and {Fouesneau}, M. and {Frenk}, C. and {Frey}, S. and
         {Gaessler}, W. and {Geier}, S. and {Fusillo}, N.~G. and {Gerhard}, O. and
         {Giannantonio}, T. and {Giannone}, D. and {Gibson}, B. and
         {Gillingham}, P. and {Gonz{\'a}lez-Fern{\'a}ndez}, C. and
         {Gonzalez-Solares}, E. and {Gottloeber}, S. and {Gould}, A. and
         {Grebel}, E.~K. and {Gueguen}, A. and {Guiglion}, G. and
         {Haehnelt}, M. and {Hahn}, T. and {Hansen}, C.~J. and {Hartman}, H. and
         {Hauptner}, K. and {Hawkins}, K. and {Haynes}, D. and {Haynes}, R. and
         {Heiter}, U. and {Helmi}, A. and {Aguayo}, C.~H. and {Hewett}, P. and
         {Hinton}, S. and {Hobbs}, D. and {Hoenig}, S. and {Hofman}, D. and
         {Hook}, I. and {Hopgood}, J. and {Hopkins}, A. and {Hourihane}, A. and
         {Howes}, L. and {Howlett}, C. and {Huet}, T. and {Irwin}, M. and
         {Iwert}, O. and {Jablonka}, P. and {Jahn}, T. and {Jahnke}, K. and
         {Jarno}, A. and {Jin}, S. and {Jofre}, P. and {Johl}, D. and
         {Jones}, D. and {J{\"o}nsson}, H. and {Jordan}, C. and
         {Karovicova}, I. and {Khalatyan}, A. and {Kelz}, A. and
         {Kennicutt}, R. and {King}, D. and {Kitaura}, F. and {Klar}, J. and
         {Klauser}, U. and {Kneib}, J. -P. and {Koch}, A. and {Koposov}, S. and
         {Kordopatis}, G. and {Korn}, A. and {Kosmalski}, J. and {Kotak}, R. and
         {Kovalev}, M. and {Kreckel}, K. and {Kripak}, Y. and {Krumpe}, M. and
         {Kuijken}, K. and {Kunder}, A. and {Kushniruk}, I. and {Lam}, M.~I. and
         {Lamer}, G. and {Laurent}, F. and {Lawrence}, J. and {Lehmitz}, M. and
         {Lemasle}, B. and {Lewis}, J. and {Li}, B. and {Lidman}, C. and
         {Lind}, K. and {Liske}, J. and {Lizon}, J. -L. and {Loveday}, J. and
         {Ludwig}, H. -G. and {McDermid}, R.~M. and {Maguire}, K. and
         {Mainieri}, V. and {Mali}, S. and {Mandel}, H. and {Mandel}, K. and
         {Mannering}, L. and {Martell}, S. and {Martinez Delgado}, D. and
         {Matijevic}, G. and {McGregor}, H. and {McMahon}, R. and
         {McMillan}, P. and {Mena}, O. and {Merloni}, A. and {Meyer}, M.~J. and
         {Michel}, C. and {Micheva}, G. and {Migniau}, J. -E. and {Minchev}, I. and
         {Monari}, G. and {Muller}, R. and {Murphy}, D. and {Muthukrishna}, D. and
         {Nandra}, K. and {Navarro}, R. and {Ness}, M. and {Nichani}, V. and
         {Nichol}, R. and {Nicklas}, H. and {Niederhofer}, F. and {Norberg}, P. and
         {Obreschkow}, D. and {Oliver}, S. and {Owers}, M. and {Pai}, N. and
         {Pankratow}, S. and {Parkinson}, D. and {Paschke}, J. and
         {Paterson}, R. and {Pecontal}, A. and {Parry}, I. and {Phillips}, D. and
         {Pillepich}, A. and {Pinard}, L. and {Pirard}, J. and {Piskunov}, N. and
         {Plank}, V. and {Pl{\"u}schke}, D. and {Pons}, E. and {Popesso}, P. and
         {Power}, C. and {Pragt}, J. and {Pramskiy}, A. and {Pryer}, D. and
         {Quattri}, M. and {Queiroz}, A.~B. d. A. and {Quirrenbach}, A. and
         {Rahurkar}, S. and {Raichoor}, A. and {Ramstedt}, S. and {Rau}, A. and
         {Recio-Blanco}, A. and {Reiss}, R. and {Renaud}, F. and {Revaz}, Y. and
         {Rhode}, P. and {Richard}, J. and {Richter}, A.~D. and {Rix}, H. -W. and
         {Robotham}, A.~S.~G. and {Roelfsema}, R. and {Romaniello}, M. and
         {Rosario}, D. and {Rothmaier}, F. and {Roukema}, B. and {Ruchti}, G. and
         {Rupprecht}, G. and {Rybizki}, J. and {Ryde}, N. and {Saar}, A. and
         {Sadler}, E. and {Sahl{\'e}n}, M. and {Salvato}, M. and {Sassolas}, B. and
         {Saunders}, W. and {Saviauk}, A. and {Sbordone}, L. and {Schmidt}, T. and
         {Schnurr}, O. and {Scholz}, R. -D. and {Schwope}, A. and {Seifert}, W. and
         {Shanks}, T. and {Sheinis}, A. and {Sivov}, T. and
         {Sk{\'u}lad{\'o}ttir}, {\'A}. and {Smartt}, S. and {Smedley}, S. and
         {Smith}, G. and {Smith}, R. and {Sorce}, J. and {Spitler}, L. and
         {Starkenburg}, E. and {Steinmetz}, M. and {Stilz}, I. and {Storm}, J. and
         {Sullivan}, M. and {Sutherland}, W. and {Swann}, E. and {Tamone}, A. and
         {Taylor}, E.~N. and {Teillon}, J. and {Tempel}, E. and {ter Horst}, R. and
         {Thi}, W. -F. and {Tolstoy}, E. and {Trager}, S. and {Traven}, G. and
         {Tremblay}, P. -E. and {Tresse}, L. and {Valentini}, M. and
         {van de Weygaert}, R. and {van den Ancker}, M. and {Veljanoski}, J. and
         {Venkatesan}, S. and {Wagner}, L. and {Wagner}, K. and
         {Walcher}, C.~J. and {Waller}, L. and {Walton}, N. and {Wang}, L. and
         {Winkler}, R. and {Wisotzki}, L. and {Worley}, C.~C. and {Worseck}, G. and
         {Xiang}, M. and {Xu}, W. and {Yong}, D. and {Zhao}, C. and {Zheng}, J. and
         {Zscheyge}, F. and {Zucker}, D.},
        title = "{4MOST: Project overview and information for the First Call for Proposals}",
      journal = {The Messenger},
         year = "2019",
        month = "Mar",
       volume = {175},
        pages = {3-11},
          doi = {10.18727/0722-6691/5117},
archivePrefix = {arXiv},
       eprint = {1903.02464},
 primaryClass = {astro-ph.IM},
       adsurl = {https://ui.adsabs.harvard.edu/abs/2019Msngr.175....3D}
}

@article{_scikit,
 title = {scikit-image: image processing in {P}ython},
 author = {van der Walt, {S}t\'efan and {S}ch\"onberger, {J}ohannes {L}. and
           {Nunez-Iglesias}, {J}uan and {B}oulogne, {F}ran\c{c}ois and {W}arner,
           {J}oshua {D}. and {Y}ager, {N}eil and {G}ouillart, {E}mmanuelle and
           {Y}u, {T}ony and the scikit-image contributors},
 year = {2014},
 month = {6},
 volume = {2},
 pages = {e453},
 journal = {PeerJ},
 issn = {2167-8359},
 url = {https://doi.org/10.7717/peerj.453},
 doi = {10.7717/peerj.453}
}

@ARTICLE{_scikit_learn,
       author = {{Pedregosa}, Fabian and {Varoquaux}, Ga{\"e}l and {Gramfort}, Alexandre and {Michel}, Vincent and {Thirion}, Bertrand and {Grisel}, Olivier and {Blondel}, Mathieu and {M{\"u}ller}, Andreas and {Nothman}, Joel and {Louppe}, Gilles and {Prettenhofer}, Peter and {Weiss}, Ron and {Dubourg}, Vincent and {Vanderplas}, Jake and {Passos}, Alexandre and {Cournapeau}, David and {Brucher}, Matthieu and {Perrot}, Matthieu and {Duchesnay}, {\'E}douard},
        title = "{Scikit-learn: Machine Learning in Python}",
      journal = {Journal of Machine Learning Research},
         year = 2011,
        month = oct,
       volume = {12},
        pages = {2825-2830},
          doi = {10.48550/arXiv.1201.0490},
archivePrefix = {arXiv},
       eprint = {1201.0490},
 primaryClass = {cs.LG},
       adsurl = {https://ui.adsabs.harvard.edu/abs/2011JMLR...12.2825P}
}

@ARTICLE{_lane23,
       author = {{Lane}, James M.~M. and {Bovy}, Jo and {Mackereth}, J. Ted},
        title = "{The stellar mass of the Gaia-Sausage/Enceladus accretion remnant}",
      journal = {\mnras},
         year = 2023,
        month = nov,
       volume = {526},
       number = {1},
        pages = {1209-1234},
          doi = {10.1093/mnras/stad2834},
archivePrefix = {arXiv},
       eprint = {2306.03084},
 primaryClass = {astro-ph.GA},
       adsurl = {https://ui.adsabs.harvard.edu/abs/2023MNRAS.526.1209L}
}

@ARTICLE{_matsuno21,
       author = {{Matsuno}, Tadafumi and {Hirai}, Yutaka and {Tarumi}, Yuta and {Hotokezaka}, Kenta and {Tanaka}, Masaomi and {Helmi}, Amina},
        title = "{R-process enhancements of Gaia-Enceladus in GALAH DR3}",
      journal = {\aap},
         year = 2021,
        month = jun,
       volume = {650},
          eid = {A110},
        pages = {A110},
          doi = {10.1051/0004-6361/202040227},
archivePrefix = {arXiv},
       eprint = {2101.07791},
 primaryClass = {astro-ph.GA},
       adsurl = {https://ui.adsabs.harvard.edu/abs/2021A&A...650A.110M}
}

@ARTICLE{_horta23,
       author = {{Horta}, Danny and {Schiavon}, Ricardo P. and {Mackereth}, J. Ted and {Weinberg}, David H. and {Hasselquist}, Sten and {Feuillet}, Diane and {O'Connell}, Robert W. and {Anguiano}, Borja and {Allende-Prieto}, Carlos and {Beaton}, Rachael L. and {Bizyaev}, Dmitry and {Cunha}, Katia and {Geisler}, Doug and {Garc{\'\i}a-Hern{\'a}ndez}, D.~A. and {Holtzman}, Jon and {J{\"o}nsson}, Henrik and {Lane}, Richard R. and {Majewski}, Steve R. and {M{\'e}sz{\'a}ros}, Szabolcs and {Minniti}, Dante and {Nitschelm}, Christian and {Shetrone}, Matthew and {Smith}, Verne V. and {Zasowski}, Gail},
        title = "{The chemical characterization of halo substructure in the Milky Way based on APOGEE}",
      journal = {\mnras},
         year = 2023,
        month = apr,
       volume = {520},
       number = {4},
        pages = {5671-5711},
          doi = {10.1093/mnras/stac3179},
archivePrefix = {arXiv},
       eprint = {2204.04233},
 primaryClass = {astro-ph.GA},
       adsurl = {https://ui.adsabs.harvard.edu/abs/2023MNRAS.520.5671H}
}

@ARTICLE{_simpson21,
       author = {{Simpson}, Jeffrey D. and {Martell}, Sarah L. and {Buder}, Sven and {Bland-Hawthorn}, Joss and {Casey}, Andrew R. and {de Silva}, Gayandhi M. and {D'Orazi}, Valentina and {Freeman}, Ken C. and {Hayden}, Michael and {Kos}, Janez and {Lewis}, Geraint F. and {Lind}, Karin and {Schlesinger}, Katharine J. and {Sharma}, Sanjib and {Stello}, Dennis and {Zucker}, Daniel B. and {Zwitter}, Toma{\v{z}} and {Asplund}, Martin and {da Costa}, Gary and {{\v{C}}otar}, Klemen and {Tepper-Garc{\'\i}a}, Thor and {Horner}, Jonathan and {Nordlander}, Thomas and {Ting}, Yuan-Sen and {Wyse}, Rosemary F.~G. and {Galah Collaboration}},
        title = "{The GALAH survey: accreted stars also inhabit the Spite plateau}",
      journal = {\mnras},
         year = 2021,
        month = oct,
       volume = {507},
       number = {1},
        pages = {43-54},
          doi = {10.1093/mnras/stab2012},
archivePrefix = {arXiv},
       eprint = {2011.02659},
 primaryClass = {astro-ph.GA},
       adsurl = {https://ui.adsabs.harvard.edu/abs/2021MNRAS.507...43S}
}

@ARTICLE{_molaro20,
       author = {{Molaro}, P. and {Cescutti}, G. and {Fu}, X.},
        title = "{Lithium and beryllium in the Gaia-Enceladus galaxy}",
      journal = {\mnras},
         year = 2020,
        month = aug,
       volume = {496},
       number = {3},
        pages = {2902-2909},
          doi = {10.1093/mnras/staa1653},
archivePrefix = {arXiv},
       eprint = {2006.00787},
 primaryClass = {astro-ph.GA},
       adsurl = {https://ui.adsabs.harvard.edu/abs/2020MNRAS.496.2902M}
}

@ARTICLE{_z0,
       author = {{Bennett}, Morgan and {Bovy}, Jo},
        title = "{Vertical waves in the solar neighbourhood in Gaia DR2}",
      journal = {\mnras},
         year = 2019,
        month = jan,
       volume = {482},
       number = {1},
        pages = {1417-1425},
          doi = {10.1093/mnras/sty2813},
archivePrefix = {arXiv},
       eprint = {1809.03507},
 primaryClass = {astro-ph.GA},
       adsurl = {https://ui.adsabs.harvard.edu/abs/2019MNRAS.482.1417B}
}

@ARTICLE{_tolstoy09,
       author = {{Tolstoy}, Eline and {Hill}, Vanessa and {Tosi}, Monica},
        title = "{Star-Formation Histories, Abundances, and Kinematics of Dwarf Galaxies in the Local Group}",
      journal = {\araa},
         year = 2009,
        month = sep,
       volume = {47},
       number = {1},
        pages = {371-425},
          doi = {10.1146/annurev-astro-082708-101650},
archivePrefix = {arXiv},
       eprint = {0904.4505},
 primaryClass = {astro-ph.CO},
       adsurl = {https://ui.adsabs.harvard.edu/abs/2009ARA&A..47..371T}
}

@ARTICLE{_helmi18,
       author = {{Helmi}, Amina and {Babusiaux}, Carine and {Koppelman}, Helmer H. and
        {Massari}, Davide and {Veljanoski}, Jovan and {Brown}, Anthony
        G.~A.},
        title = "{The merger that led to the formation of the Milky Way's inner stellar halo and thick disk}",
      journal = {\nat},
         year = 2018,
        month = Nov,
       volume = {563},
        pages = {85-88},
          doi = {10.1038/s41586-018-0625-x},
archivePrefix = {arXiv},
       eprint = {1806.06038},
 primaryClass = {astro-ph.GA},
       adsurl = {https://ui.adsabs.harvard.edu/\#abs/2018Natur.563...85H}
}

@ARTICLE{_trick18,
       author = {{Trick}, Wilma H. and {Coronado}, Johanna and {Rix}, Hans-Walter},
        title = "{The Galactic disc in action space as seen by Gaia DR2}",
      journal = {\mnras},
         year = "2019",
        month = "Apr",
       volume = {484},
       number = {3},
        pages = {3291-3306},
          doi = {10.1093/mnras/stz209},
archivePrefix = {arXiv},
       eprint = {1805.03653},
 primaryClass = {astro-ph.GA},
       adsurl = {https://ui.adsabs.harvard.edu/abs/2019MNRAS.484.3291T}
}

@ARTICLE{_koppelman18,
       author = {{Koppelman}, Helmer and {Helmi}, Amina and {Veljanoski}, Jovan},
        title = "{One Large Blob and Many Streams Frosting the nearby Stellar Halo in Gaia
        DR2}",
      journal = {\apj},
         year = 2018,
        month = Jun,
       volume = {860},
          eid = {L11},
        pages = {L11},
          doi = {10.3847/2041-8213/aac882},
 primaryClass = {astro-ph.GA},
       adsurl = {https://ui.adsabs.harvard.edu/#abs/2018ApJ...860L..11K}
}

@ARTICLE{_ramos18,
       author = {{Ramos}, P. and {Antoja}, T. and {Figueras}, F.},
        title = "{Riding the kinematic waves in the Milky Way disk with Gaia}",
      journal = {\aap},
         year = "2018",
        month = "Nov",
       volume = {619},
          eid = {A72},
        pages = {A72},
          doi = {10.1051/0004-6361/201833494},
archivePrefix = {arXiv},
       eprint = {1805.09790},
 primaryClass = {astro-ph.GA},
       adsurl = {https://ui.adsabs.harvard.edu/abs/2018A&A...619A..72R}
}

@ARTICLE{_kushniruk17,
       author = {{Kushniruk}, I. and {Schirmer}, T. and {Bensby}, T.},
        title = "{Kinematic structures of the solar neighbourhood revealed by Gaia
        DR1/TGAS and RAVE}",
      journal = {\aap},
         year = 2017,
        month = Dec,
       volume = {608},
          eid = {A73},
        pages = {A73},
          doi = {10.1051/0004-6361/201731147},
       adsurl = {https://ui.adsabs.harvard.edu/#abs/2017A&A...608A..73K}
}

@ARTICLE{_zhao12,
   author = {{Zhao}, G. and {Zhao}, Y.-H. and {Chu}, Y.-Q. and {Jing}, Y.-P. and 
	{Deng}, L.-C.},
    title = "{LAMOST spectral survey {\mdash} An overview}",
  journal = {Research in Astronomy and Astrophysics},
     year = 2012,
    month = jul,
   volume = 12,
    pages = {723-734},
      doi = {10.1088/1674-4527/12/7/002},
   adsurl = {http://adsabs.harvard.edu/abs/2012RAA....12..723Z}
}

@ARTICLE{_freeman02,
   author = {{Freeman}, K. and {Bland-Hawthorn}, J.},
    title = "{The New Galaxy: Signatures of Its Formation}",
  journal = {\araa},
   eprint = {astro-ph/0208106},
     year = 2002,
   volume = 40,
    pages = {487-537},
      doi = {10.1146/annurev.astro.40.060401.093840},
   adsurl = {http://adsabs.harvard.edu/abs/2002ARA%26A..40..487F}
}

@ARTICLE{Hinton2002,
       author = {{Hinton}, G.E. and {Roweis}, S.T.},
        title = "{Stochastic Neighbor Embedding}",
      journal = {Advances in Neural Processing Systems},
         year = 2002,
       volume = {15},
        pages = {833},
}
\end{document}